%% file: Main-Els.tex
\documentclass[preprint,number,longtitle]{elsarticle}

\usepackage{amsmath,amssymb}

\begin{document}

\begin{frontmatter}

\title{The Design and Performance of IceCube DeepCore}

\input{AuthorList-Elsart-Summer2011.tex}

\begin{abstract}
   \input{Abstract.tex}

\end{abstract}

\end{frontmatter}

\input{NewCommands.tex}

\section{Introduction}
\label{sec:Introduction}
\input{Introduction.tex}

\section{DeepCore Design and Schedule}
\label{sec:Design}

\subsection{Ice Properties}
\label{subsec:IceProperties}
\input{IceProperties.tex}

\subsection{Photomultiplier Tubes}
\label{subsec:New_Photocathode_PMTs}
\input{New_Photocathode_PMTs.tex}

\subsection{Geometry}
\label{subsec:Geometry}
\input{Geometry.tex}

\section{Simulation and Selection of DeepCore Events}
\label{sec:Simulation}

\subsection{Simulation Tools}
\label{subsec:Simulation_Tools}
\input{Simulation_Tools.tex}

\subsection{Trigger}
\label{subsec:Triggering}
\input{Triggering.tex}

\subsection{Online Filter}
\label{subsec:Filtering}
\input{Filtering.tex}

\section{Conclusions}
\label{sec:Conclusions}
\input{Conclusions.tex}

\clearpage

\section{Acknowledgements}

\input{Acknowledgements}

\clearpage

\bibliography{biblio}  

\end{document}

%% file: AuthorList-Elsart-Summer2011.tex
\author[Madison]{R.~Abbasi}
\author[Gent]{Y.~Abdou}
\author[RiverFalls]{T.~Abu-Zayyad}
\author[Zeuthen]{M.~Ackermann}
\author[Christchurch]{J.~Adams}
\author[Madison]{J.~A.~Aguilar}
\author[Oxford]{M.~Ahlers}
\author[PennPhys]{M.~M.~Allen}
\author[Aachen]{D.~Altmann}
\author[Madison]{K.~Andeen\fnref{Rutgers}}
% ...
\address[Aachen]{III. Physikalisches Institut, RWTH Aachen University, D-52056 Aachen, Germany}
% ...
\address[Christchurch]{Dept.~of Physics and Astronomy, University of Canterbury, Private Bag 4800, Christchurch, New Zealand}
% ...
\address[Gent]{Dept.~of Physics and Astronomy, University of Gent, B-9000 Gent, Belgium}
% ...
\address[Madison]{Dept.~of Physics, University of Wisconsin, Madison, WI 53706, USA}
% ...
\address[Oxford]{Dept.~of Physics, University of Oxford, 1 Keble Road, Oxford OX1 3NP, UK}
\address[RiverFalls]{Dept.~of Physics, University of Wisconsin, River Falls, WI 54022, USA}
% ...
\address[PennPhys]{Dept.~of Physics, Pennsylvania State University, University Park, PA 16802, USA}
% ...
\address[Zeuthen]{DESY, D-15735 Zeuthen, Germany}
\fntext[Rutgers]{Now at Dept. of Physics and Astronomy, Rutgers University, Piscataway, NJ 08854, USA}
\fntext[SouthDakota]{Now at Physics Department, South Dakota School of Mines and Technology, Rapid City, SD 57701, USA}
\fntext[LosAlamos]{Los Alamos National Laboratory, Los Alamos, NM 87545, USA}
\fntext[Bari]{Also Sezione INFN, Dipartimento di Fisica, I-70126, Bari, Italy}
\fntext[TumErl]{Now at T.U. Munich, 85748 Garching \& Friedrich-Alexander Universit\"at Erlangen-N\"urnberg, 91058 Erlangen, Germany}
\fntext[TUM]{Now at T.U. Munich, 85748 Garching, Germany}
\fntext[Goddard]{NASA Goddard Space Flight Center, Greenbelt, MD 20771, USA}

%% file: Abstract.tex
The IceCube neutrino observatory in operation at the South Pole,
Antarctica, comprises three distinct components: a large buried array
for ultrahigh energy neutrino detection, a surface air shower array,
and a new buried component called DeepCore.  DeepCore was designed to
lower the IceCube neutrino energy threshold by over an order of
magnitude, to energies as low as about 10~GeV.  DeepCore is situated
primarily 2100~m below the surface of the icecap at the South Pole, at
the bottom center of the existing IceCube array, and began taking
physics data in May 2010.  Its location takes advantage of the
exceptionally clear ice at those depths and allows it to use the
surrounding IceCube detector as a highly efficient active veto against
the principal background of downward-going muons produced in
cosmic-ray air showers. DeepCore has a module density roughly five
times higher than that of the standard IceCube array, and uses
photomultiplier tubes with a new photocathode featuring a quantum
efficiency about 35\% higher than standard IceCube PMTs.  Taken
together, these features of DeepCore will increase IceCube's
sensitivity to neutrinos from WIMP dark matter annihilations,
atmospheric neutrino oscillations, galactic supernova neutrinos, and
point sources of neutrinos in the northern and southern skies.  In
this paper we describe the design and initial performance of DeepCore.

%% file: NewCommands.tex
\newcommand{\gsim}{\gtrsim}
\newcommand{\lsim}{\lesssim}
\newcommand{\Enu}{\rm{E}_\nu}
\newcommand{\nue}{\nu_{\rm e}}
\newcommand{\numu}{\nu_\mu}
\newcommand{\nutau}{\nu_\tau}

\newcommand{\Nch}{${\rm N}_{\rm ch}\,$}
\newcommand{\Ndir}{${\rm N}_{\rm dir}\,$}
\newcommand{\Nstr}{${\rm N}_{\rm str}\,$}
\newcommand{\Aeff}{${\rm A}_{\rm eff}\,$}
\newcommand{\Veff}{${\rm V}_{\rm eff}\,$}
\newcommand{\VeffNS}{${\rm V}_{\rm eff}$}

%% file: Introduction.tex
DeepCore is a new subarray of the IceCube observatory~\cite{IceCube}
that was designed to provide sensitivity to neutrinos at energies over
an order of magnitude lower than initially envisioned for the original
array.  Using the Cherenkov light emitted by charged particles arising
from neutrino interactions in the ice, the subarray achieves this
improved sensitivity through a combination of increased module
density, higher quantum efficiency photomultiplier tubes (PMTs),
deployment in the clearest ice at depths greater than 2100~m, and use
of the surrounding standard IceCube modules above and around DeepCore
as a powerful active veto against the copious downward-going
cosmic-ray muon background.

DeepCore provides enhanced sensitivity to weakly interacting massive
particles (WIMPs) and is expected to significantly improve existing
IceCube results on WIMP annihilations in the Sun~\cite{IC22WIMPs},
Galactic Center~\cite{WIMPs_GC} and Halo~\cite{WIMPS_GH}, extending
limits below present accelerator bounds.  DeepCore gives improved
acceptance for low energy atmospheric neutrinos at $\Enu \gsim
10$~GeV, opening a useful new window for atmospheric neutrino
oscillation measurements, including $\numu$ disappearance, $\nutau$
appearance and as a remote possibility, the sign of the neutrino
hierarchy~\cite{Mena:2008rh}.  Taking advantage of the active vetoing
capability provided by the surrounding IceCube array, DeepCore allows
us to explore the southern sky for diffuse and point source neutrino
emission from active galactic nuclei (AGN), gamma ray bursts (GRBs),
choked GRBs~\cite{2010PhRvD81h3011T}, and the inner galaxy.  The
increased module density of DeepCore may enable the reconstruction of
more closely-spaced cascades produced by an initial $\nutau$
interaction and the subsequent $\tau$ decay, extending the search for
cosmological $\nutau$ to lower energies.  The higher module density
may also enable the reconstruction of the average energy of galactic
supernova neutrinos~\cite{SN_ICRC,SN_ArXiv}.  Searches for slow-moving
monopoles, supersymmetric stau pair
production~\cite{staus,SUSYKaluzaKleinMultipleScattering} and
low-energy neutrino emission from astrophysical
sources~\cite{LowEGRBNus} will likewise benefit from DeepCore's
extension of IceCube's capabilities.

Section~\ref{sec:Design} of this paper describes the design of
DeepCore, highlighting the geometrical layout of the sub-array of
digital optical modules (DOMs) that house the PMTs and their
associated readout electronics~\cite{IceCubeDOMs,IceCubeDAQ}, the
optical quality of the ice in which DeepCore has been deployed, the
performance characteristics of its high quantum efficiency PMTs, and
the schedule for DeepCore deployment that led to ``first light'' in
mid-2010.  In Section~\ref{sec:Simulation} we describe the results of
simulations performed with IceCube and DeepCore, showing predicted
triggering and quasi-real-time event selection (``filtering'')
performance, and estimations of neutrino effective volumes.
Section~\ref{sec:Conclusions} gives our conclusions.

%% file: IceProperties.tex
The Antarctic Muon and Neutrino Detector Array (AMANDA, the
predecessor to IceCube) was used to map~\cite{icepaper} the relevant
wavelength and depth dependence of light absorption and scattering
down to a depth of 2350~m, albeit with poor precision beyond 2100~m
because of much sparser instrumentation in the deepest ice. Variations
with depth in the optical ice properties were found to closely track
concentrations in mineral dust deposits which in turn are correlated
with climatological history. Dust concentrations are the highest in a
depth band between 2000 and 2100~m, here denoted the ``dust layer,''
corresponding to a stadial about 65,000 years ago, in the last glacial
period.  DeepCore was designed to avoid this highly absorbing and
scattering ice.

Our knowledge of the optical properties of the ice in which DeepCore
is located has been augmented more recently with {\it in situ}
measurements using pulsed LED sources in IceCube. These preliminary
time-of-flight measurements verify that the ice at depths greater than
2100~m is significantly more transparent optically than the shallower
ice between 1500 and 2000~m. We also see this qualitatively in IceCube
data from downward-going muons, which show a strong light depletion in
the dust layer followed by increased light yield at greater depths
(see Fig.~\ref{fig:HQEPMTOccupancy}).

In terms of scattering and absorption lengths, the parameters
describing photon propagation that need to be known to simulate (see
Sec.~\ref{subsec:Simulation_Tools}) and reconstruct neutrino-induced
events, the average ice at depths below 2100~m is estimated to be
about 40\%-50\% clearer than the ice between 1500 and 2000~m. In the
clearest ice, around 2400~m depth, the average effective scattering
length is close to 50~m and the average absorption length is close to
190~m. These values are for 400~nm light, the wavelength where
absorption due to dust is weakest and the ice is most transparent.
This wavelength is also near the peak of the DOM
sensitivity~\cite{IceCubeDOMs}.

%% file: New_Photocathode_PMTs.tex
\begin{figure} [h!]
   \begin{center}
      \includegraphics[width=7cm, angle=0]{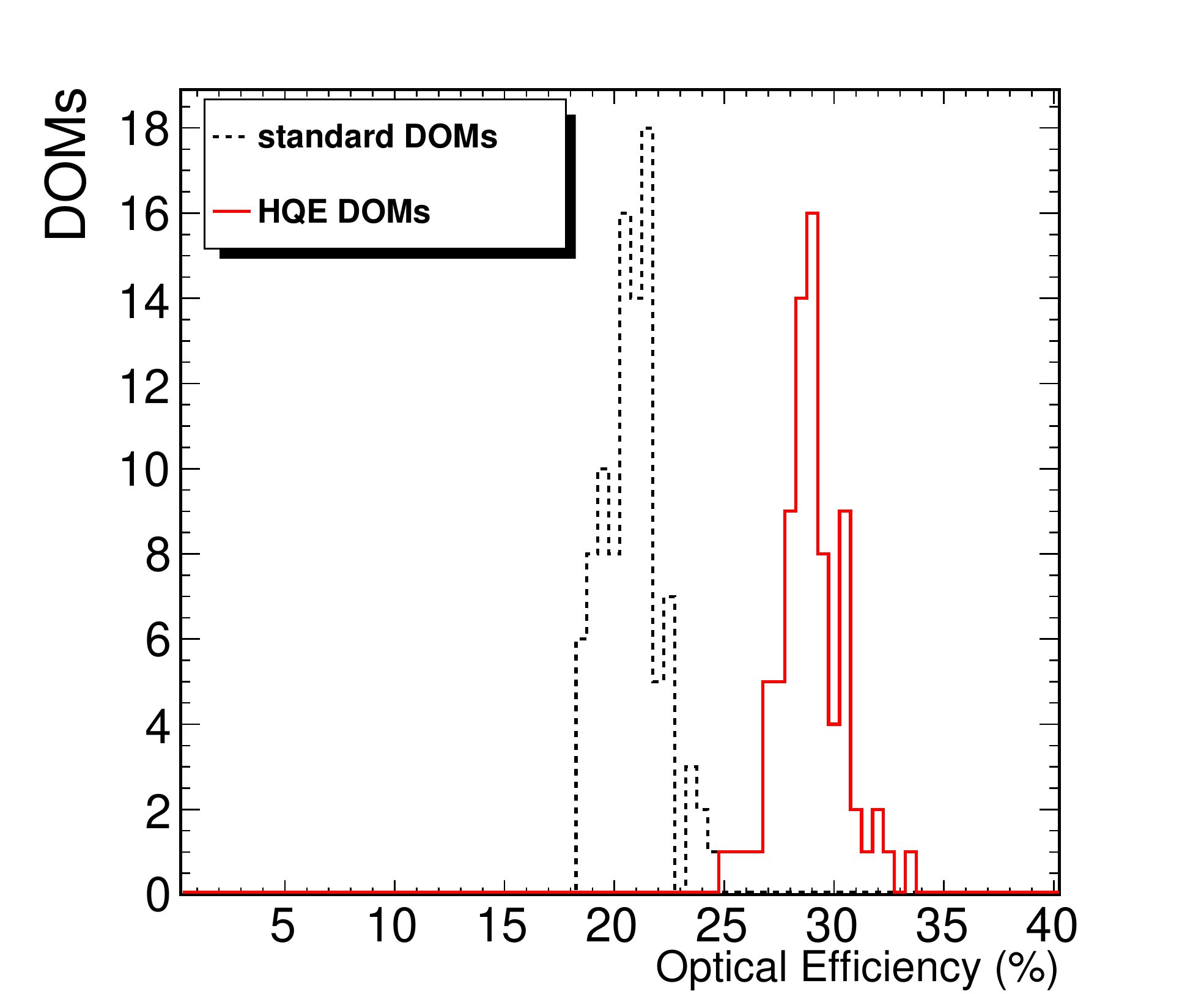}
   \end{center}
   \caption{Laboratory measurement of the relative optical
     efficiencies at $-45^\circ{\rm C}$ and 405~nm, near the peak PMT
     sensitivity.  A standard reference 2-inch PMT was used for
     normalization purposes.  The black dashed curve is for DOMs with
     standard PMTs, the red solid curve for HQE DOMs.  Using the ratio
     of the mean relative efficiencies, the new PMTs have an
     efficiency 1.39 times higher than the standard PMTs. }
   \label{fig:HQEPMTEfficiency}
\end{figure}
\begin{figure} [h!]
   \begin{center}
      \includegraphics[width=7cm, angle=0]{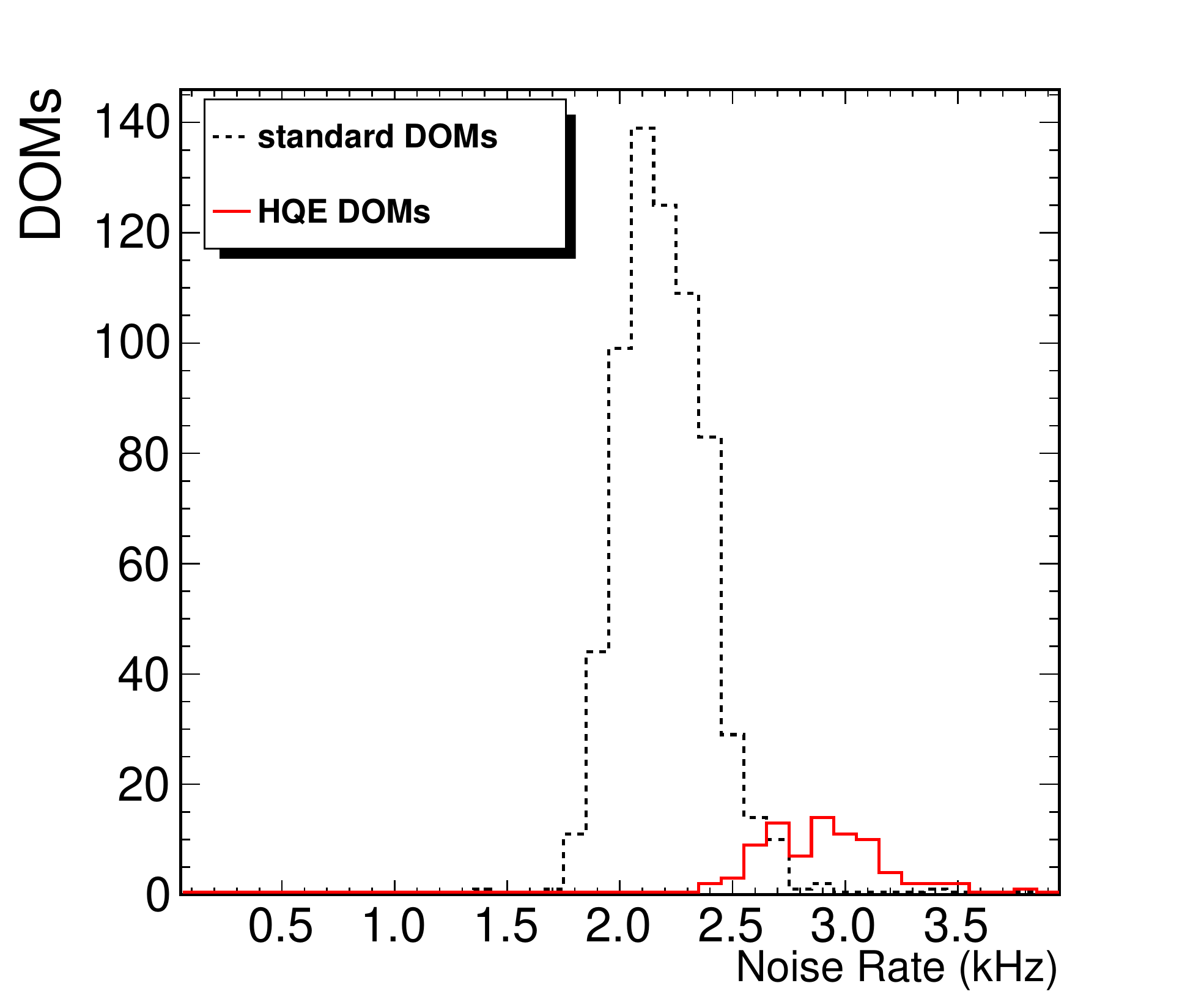}
   \end{center}
   \caption{Laboratory measurement of the HQE DOM noise rate for the
     DOM at $-45^\circ{\rm C}$.  The black dashed curve is for DOMs with
     standard PMTs, the red solid curve for HQE DOMs.}
   \label{fig:HQEPMTNoiseFAT}
\end{figure}
\begin{figure} [h!]
   \begin{center}
      \includegraphics[width=7cm, angle=0]{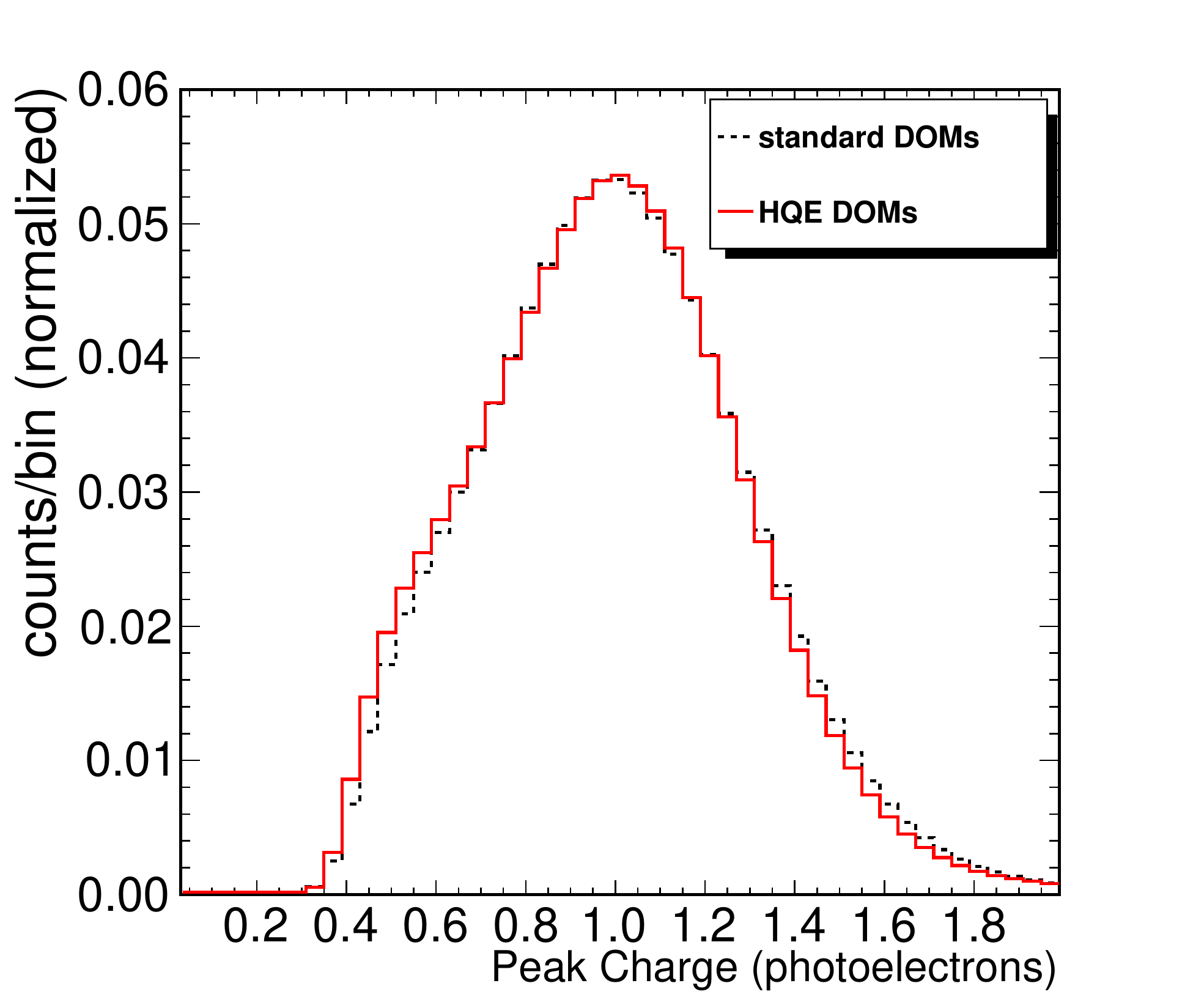}
   \end{center}
   \caption{Laboratory measurement of the HQE DOM charge response for
     the DOM at $-45^\circ{\rm C}$.  The black dashed curve is for DOMs with
     standard PMTs, the red solid curve for HQE DOMs.}
   \label{fig:HQEPMTQResponse}
\end{figure}

The photomultiplier tube used in the standard IceCube DOMs is the
Hamamatsu 252~mm diameter R7081-02~\cite{IceCubeDOMs}. During the
planning for DeepCore, Hamamatsu presented a new version of the PMT,
the R7081MOD, with higher quantum efficiency.  The R7081MOD is
identical to the standard IceCube PMT, but with a ``super bialkali''
photocathode that improves the quantum efficiency by about 40\% at
photon wavelength $\lambda = 390$~nm in laboratory measurements
performed by Hamamatsu. Eight of these new PMTs were tested in the
laboratory by IceCube, confirming the higher quantum
efficiency. Subsequent simulations demonstrated that the added
efficiency would increase the effective area of DeepCore for
triggering on low energy neutrinos by about 30\%.  All DOMs equipped
with the new high quantum efficiency PMTs (``HQE DOMs'') were fully
tested in the standard IceCube DOM testing system and the results from
the first 80 of these are compared with the standard DOMs in
Figs.~\ref{fig:HQEPMTEfficiency},~\ref{fig:HQEPMTNoiseFAT}
and~\ref{fig:HQEPMTQResponse}.  These tests included several
temperature cycles from $+25^\circ{\rm C}$ to $-45^\circ{\rm C}$ to
simulate the temperatures the DOMs would experience during
transportation and deployment.  Once deployed, the ambient temperature
varies between approximately $-45^\circ{\rm C}$ to $-20^\circ{\rm C}$,
becoming warmer with increasing depth.~\cite{IceTemperatureProfile}

\begin{figure} [h!]
   \begin{center}
      \includegraphics[width=8cm, angle=0]{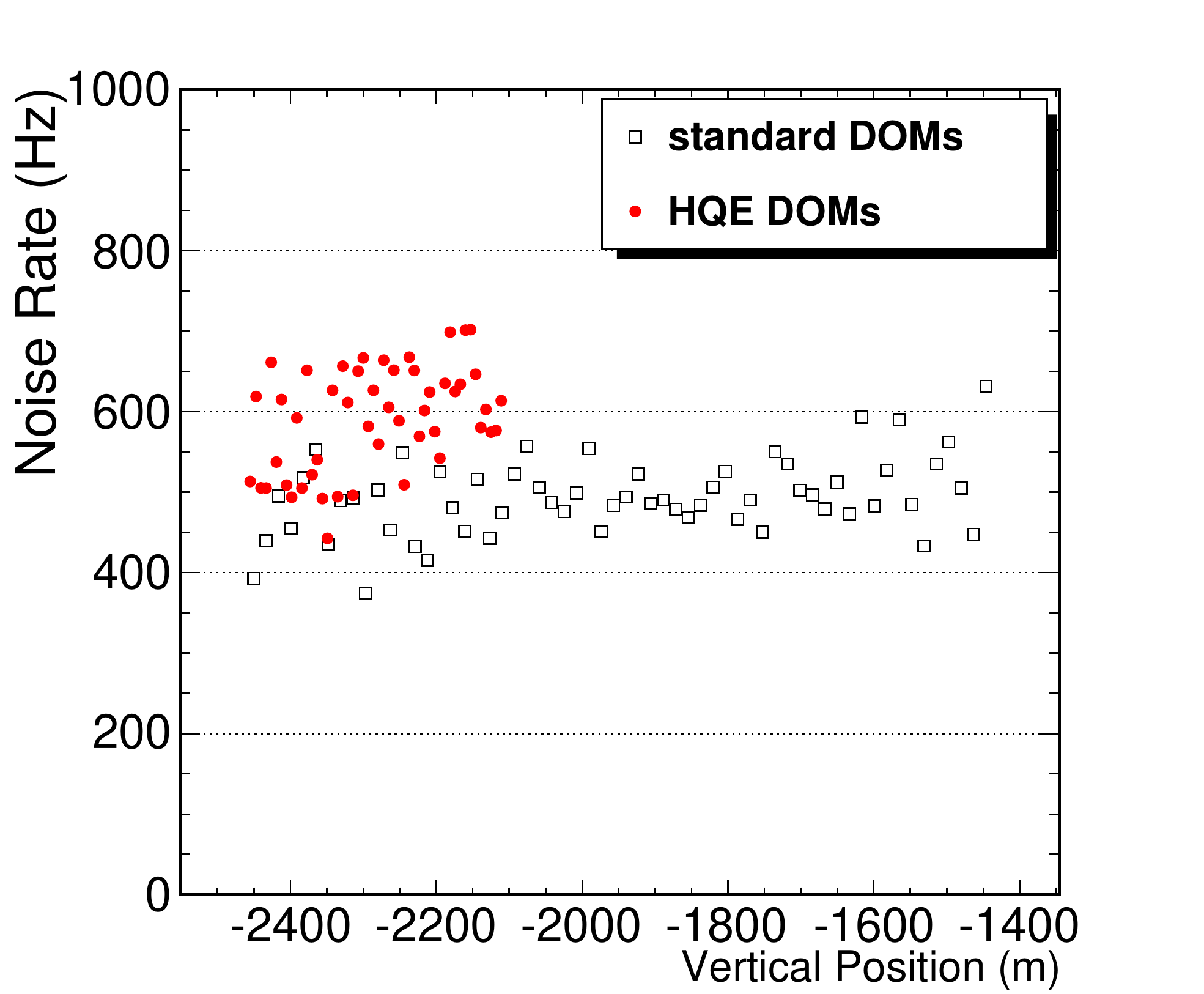}
   \end{center}
   \caption{HQE DOM noise rates from {\it in situ} measurements as a
     function of deployment depth.  The black squares are for
     DOMs with standard PMTs, the red circles for HQE DOMs.  The
     higher noise rate is consistent with the increased quantum
     efficiency.}
   \label{fig:HQEPMTNoiseHLC}
\end{figure}

\begin{figure} [h!]
   \begin{center}
      \includegraphics[width=8cm, angle=0]{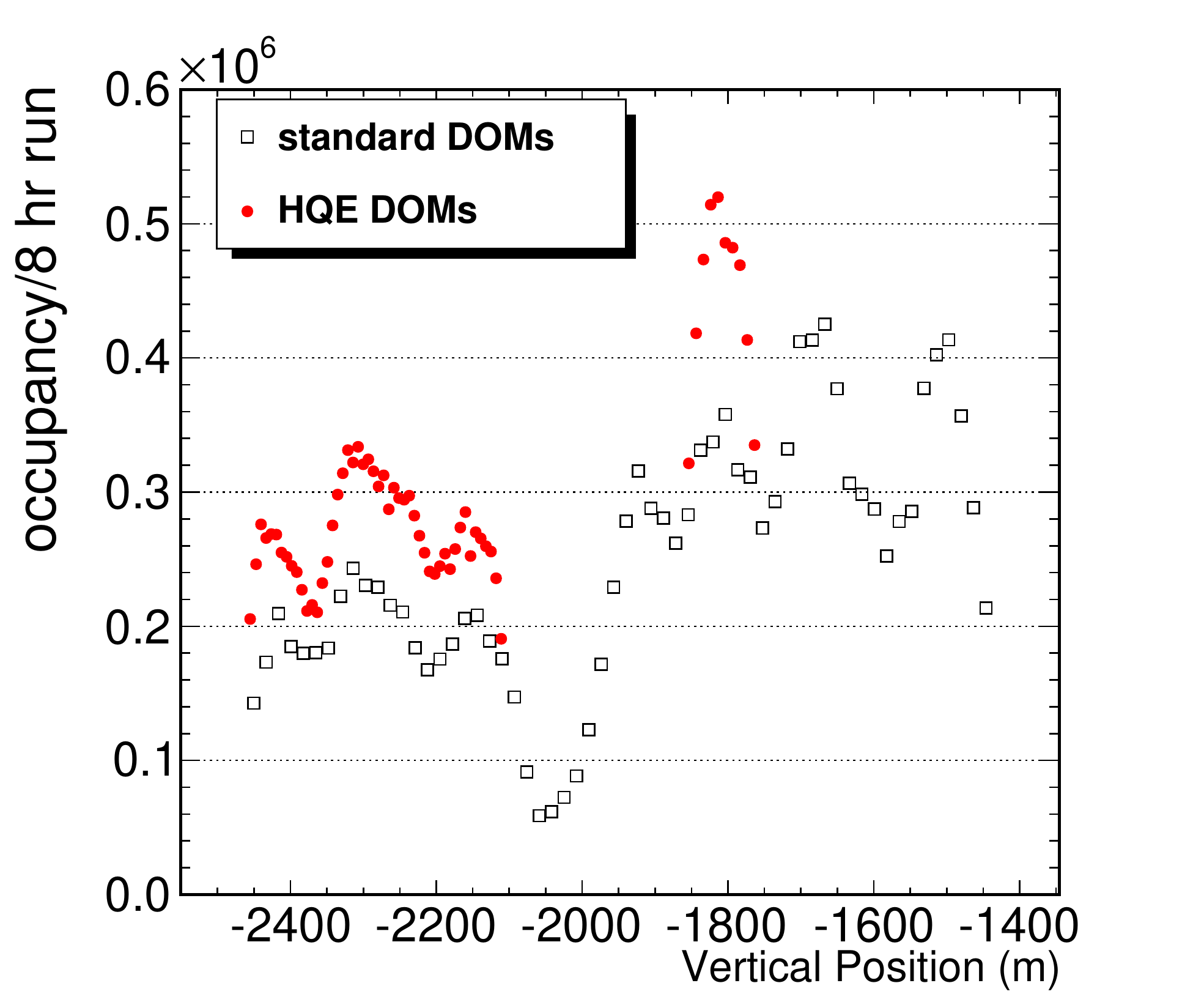}
   \end{center}
   \caption{HQE DOM occupancies (defined in the text) from {\it in
       situ} measurements as a function of deployment depth.  The
     black squares are for standard PMTs, the red circles
     are for HQE PMTs.  The latter show a higher occupancy due to
     improved quantum efficiency.}
   \label{fig:HQEPMTOccupancy}
\end{figure}

In the laboratory measurements the HQE DOMs showed a 39\% higher
optical sensitivity than standard DOMs at $\lambda=405$~nm
(Fig.~\ref{fig:HQEPMTEfficiency}).  An {\it in-situ} measurement
showed an improvement of about 35\%, smaller than the laboratory
measurements and possibly due to the non-monochromatic Cherenkov
spectrum and to the different optical system created by the
surrounding ice.  Additional properties of the HQE DOMs that differ
from those of standard DOMs include: an average noise rate that is
higher by a factor of 1.33 at $-45^\circ{\rm C}$ and with a
programmable deadtime set to $100$~ns (Fig.~\ref{fig:HQEPMTNoiseFAT});
a high voltage at $10^7$ gain that is 100~V lower; and a slightly
larger peak-to-valley ratio.  Standard and HQE DOMs exhibited similar
photo-electron pulse height and charge spectra
(Fig.~\ref{fig:HQEPMTQResponse}).  Figure~\ref{fig:HQEPMTNoiseHLC}
shows the dark noise rates and Fig.~\ref{fig:HQEPMTOccupancy} shows
the relative occupancies for standard and HQE DOMs, measured {\it in
  situ}.  The occupancy is defined as the fraction of events in which
each DOM detected one or more photons and is on a string with at least
seven other DOMs that also detected photons.  Each DOM must also be in
``hard local coincidence,'' a condition that requires it to have at
least one neighboring DOM registering a hit contemporaneously (see
Sec.~\ref{subsec:Simulation_Tools}).

%% file: Geometry.tex
The geometric parameters that directly impact the ability of DeepCore
to reconstruct low energy events and discriminate them from the
cosmic-ray muon background are its horizontal string-to-string and
vertical DOM-to-DOM spacings.  Since low energy muons ($E_\mu \sim
1$~TeV) in ice travel about 5~m per GeV of energy, the 125~m
horizontal string spacing and 17~m vertical DOM spacing of the
baseline IceCube detector translate to a minimum neutrino energy
threshold for most analyses of about 50--100~GeV, and an optimal
response at $\Enu \gsim 1$~TeV.

Low energy events are especially susceptible to background
contamination from atmospheric cosmic-ray muons.  Cosmic-ray muons
trigger IceCube at a rate approximately $10^{6}$ times higher than
atmospheric neutrino interactions in the detector.  The flux of these
background events is greater at lower energies, and the ability to
distinguish between signal and background is more challenging in
events with low light levels.  The muon background is suppressed both
by situating DeepCore at the greatest available depths and by using
the entire surrounding IceCube detector as an active veto.  Most
downward-going cosmic-ray muons are unlikely to evade detection by the
large number of surrounding DOMs before entering the DeepCore's 125~m
radius by 350~m long cylindrical fiducial volume.

The DeepCore geometry was optimized using a Monte Carlo (MC)
simulation based on the detailed simulation tools already in use by
IceCube.  The emphasis was on maximizing the detection efficiency for
fully or partially contained (i.e., starting) neutrino events in the
10--100~GeV energy range while also achieving cosmic-ray background
rejection of $10^6$ or better.  For reference, fully-contained
upward-going muons with ${\rm E}_{\mu} \simeq 10$~GeV can illuminate
about 10~DOMs in DeepCore, a number sufficient both for triggering the
detector and for applying sophisticated reconstruction algorithms.

\begin{figure} [h!]
   \begin{center}
      \includegraphics[width=8cm, angle=0]{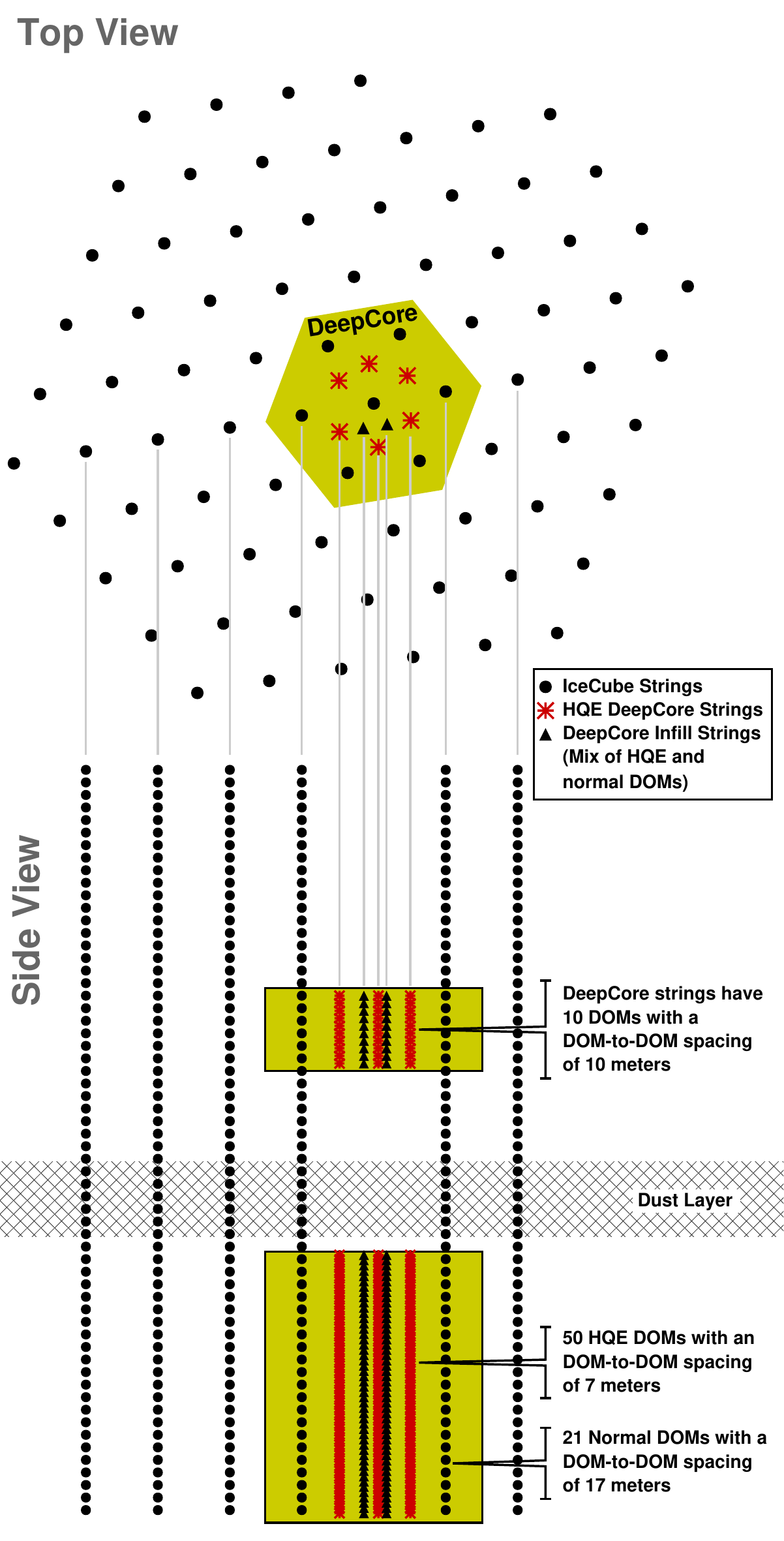}
   \end{center}
   \caption{A schematic layout of IceCube DeepCore.  The upper diagram
     shows a top view of the string positions in relation to current
     and future IceCube strings.  It includes two additional strings,
     situated close to the central DeepCore string, that were deployed
     in the 2010/2011 austral summer.  Please see the text for a
     quantitative description of the detector geometry.}
   \label{fig:DCgeometry}
\end{figure}

We varied DOM and inter-string spacing, balancing the competing
advantages of higher module density and greater fiducial volume, while
remaining consistent with drilling and down-hole cable breakout
constraints.  The chosen configuration (see Fig.~\ref{fig:DCgeometry})
comprises eight new strings, six with 60~HQE~DOMs, located very near
the bottom center of IceCube, logically joined with the bottom third
of the seven nearest-neighbor standard IceCube strings.  The average
inter-string horizontal distance between 13 of the 15 DeepCore strings
is 72~m, about 1.5 times the effective scattering length of the ice
surrounding most of DeepCore.  For six of the 15 DeepCore strings, the
interstring spacing is 42~m. On each new string, 50~DOMs with 7~m
vertical spacing are located in the deepest ice instrumented by
IceCube, between 2100 and 2450~m below the surface of the polar
icecap, where the scattering and absorption lengths are substantially
longer than at shallower depths.  The region between 2000 and 2100~m
depths is not instrumented with DeepCore DOMs due to the significant
scattering and absorption of light that occurs there.  Instead, the
remaining 10~DOMs of each of the eight high-density DeepCore strings
are placed directly above this region with a spacing of 10~m,
providing an added overhead veto ``plug'' to further enhance
background rejection from the vertical direction where the cosmic-ray
angular distribution is peaked.  These remaining DOMs also improve the
reconstruction of low-energy, near-horizontal tracks useful in certain
ongoing analyses.

The first specialized DeepCore string was successfully deployed in
January 2009~\cite{IceCubeDeployment}.  In addition, six of the
seven standard IceCube strings that are part of the DeepCore fiducial
volume were deployed.  In the following 2009/2010 austral summer
season, the five remaining DeepCore strings with HQE PMTs and the one
remaining standard IceCube string were deployed.  At that point,
DeepCore was comprised of six strings with closely spaced HQE-PMT DOMs
and seven standard IceCube strings, and was fully surrounded by three
layers of standard IceCube strings.  This DeepCore configuration
started acquiring physics data in April 2010.

The two DeepCore strings on 42~m horizontal spacing were deployed in
the 2010/2011 austral summer.  Initial MC studies have shown that this
closer-packed configuration will further improve DeepCore's low energy
response, with an estimated 30\% increased rate for events with $E_\nu
= 10-20$~GeV having six or more hits in the fiducial volume.

%% file: Simulation_Tools.tex
Monte Carlo studies have been used to design DeepCore and optimize its
geometry, to study the signal acceptance and background veto
efficiency, and to evaluate its physics potential.  The MC data for
all these studies were generated using the complete IceCube simulation
package, called IceSim.  IceSim has interfaces to various programs
needed to produce the signal and background events of interest.
Atmospheric muons were simulated with an air shower simulation program
CORSIKA (COsmic Ray SImulation KAscade)~\cite{CORSIKA} and neutrinos
were generated with code based on ANIS (All Neutrino Interaction
Generator)~\cite{ANIS} and with GENIE (Generates Events for Neutrino
Interaction Experiments)~\cite{GENIE}.  Originally developed for the
AMANDA detector and recently adapted for use in the IceCube software
framework, ANIS is capable of generating neutrinos with energies
between 10~GeV and 1~ZeV.  With a more accurate description of
neutrino interactions below $\Enu = 10$~GeV, GENIE is a
state-of-the-art generator used in the broader neutrino community, in
particular by accelerator neutrino experiments, and has been
extensively verified. It is foreseen that an extension of GENIE to
higher energies will become available in the near future, covering the
full IceCube detector energy range.

In each of these programs a parent particle was produced and
propagated to a specific boundary of the detector geometry.  For
instance, the CORSIKA-generated cosmic-ray muons were propagated to
the surface of the earth, and ANIS-generated neutrinos to a cylinder
of fixed radius around the IceCube detector.  Once the parent particle
reached the boundary, its charged lepton daughters were propagated
with MMC~\cite{MMC}, also interfaced with IceSim.

IceSim contains the full details of the IceCube detector, including
DOM hardware and firmware simulation and Photonics~\cite{PHOTONICS}
which propagates photons emitted by charged particle interactions
through the ice, taking into account local variations in its optical
properties~\cite{icepaper}.  This simulation chain, from parent
particles to the leptons and photons and finally to a DOM/PMT
simulation, produces simulated events containing the list of hit DOMs
with associated charge and timing.  The content of the simulation
output is a superset of that produced by the IceCube detector
DAQ.  The same trigger, filter and analysis algorithms are applied to
both simulated and real data.

IceSim's modular design made the inclusion of DeepCore
straightforward.  The main difference was the higher average noise
rate and improved photon detection efficiency of the HQE DOMs.  To
account for these differences in these initial studies, estimated
linear scale factors were introduced, based on preliminary lab
measurements of the HQE DOMs.  The simulated noise rate was increased by a
factor of 1.54 and the PMT quantum efficiency was increased by a
factor of 1.25 relative to standard DOMs.  (As shown in
Sec.~\ref{subsec:New_Photocathode_PMTs}, later measurements of the noise
rate and relative quantum efficiency indicate that these estimates
were approximately correct.)  Eventually, the relative quantum
efficiency for these DOMs, which is wavelength dependent, will be
included directly in our photon propagation code, once a complete
calibration of the deployed DeepCore detector has been performed.

%% file: Triggering.tex
IceCube DOMs are read out whenever a sufficient number of hits
satisfying the hard local coincidence (HLC) condition occur during a
pre-defined time window.  The HLC condition is satisfied when two or
more DOMs in close proximity to one another (nearest or next-nearest
neighbors on the same string) register hits within a $\pm 1~\mu$s time
window.  IceCube uses a simple majority trigger requiring eight or
more DOMs satisfying the HLC condition within a 5~$\mu$s time window
(this trigger is called ``SMT8'').  The detector readout is then
expanded to a wider $\pm 10~\mu$s time window centered on the trigger
time, and includes DOMs which registered hits in the trigger time
window but which did not satisfy the HLC condition.  These DOMs are
said to satisfy the ``soft local coincidence'' (SLC) condition.  Only
HLC hits are used by the trigger.  Detailed hit information is
acquired from DOMs satisfying the HLC condition as these DOMs may have
received substantial amounts of light, while less detailed information
is acquired from DOMs satisfying the SLC condition as these DOMs
typically receive only single photons~\cite{IceCubeDAQ}.

To reach lower energies, DeepCore uses an independent SMT3 trigger,
with a 2.5~$\mu$s time window, applied to DOMs comprising the DeepCore
fiducial volume--the DOMs on the strings with HQE DOMs and those on
neighboring standard IceCube strings below 2100~m.  The background to
this trigger from coincident random noise is greatly reduced by
application of the HLC condition.  Furthermore, the depth of the DOMs
suppresses the cosmic-ray background to levels manageable for the DAQ
system.  The DeepCore SMT3 trigger has an exclusive trigger rate that
is $<10$~Hz, which is $< 0.4$\% that of the standard IceCube SMT8
trigger.  Table~\ref{tab:FilterPerformance} shows the measured and
simulated SMT3 trigger rates from cosmic-rays and atmospheric
neutrinos.

\begin{figure} [h!]
   \begin{center}
      \includegraphics[width=8cm, angle=0]{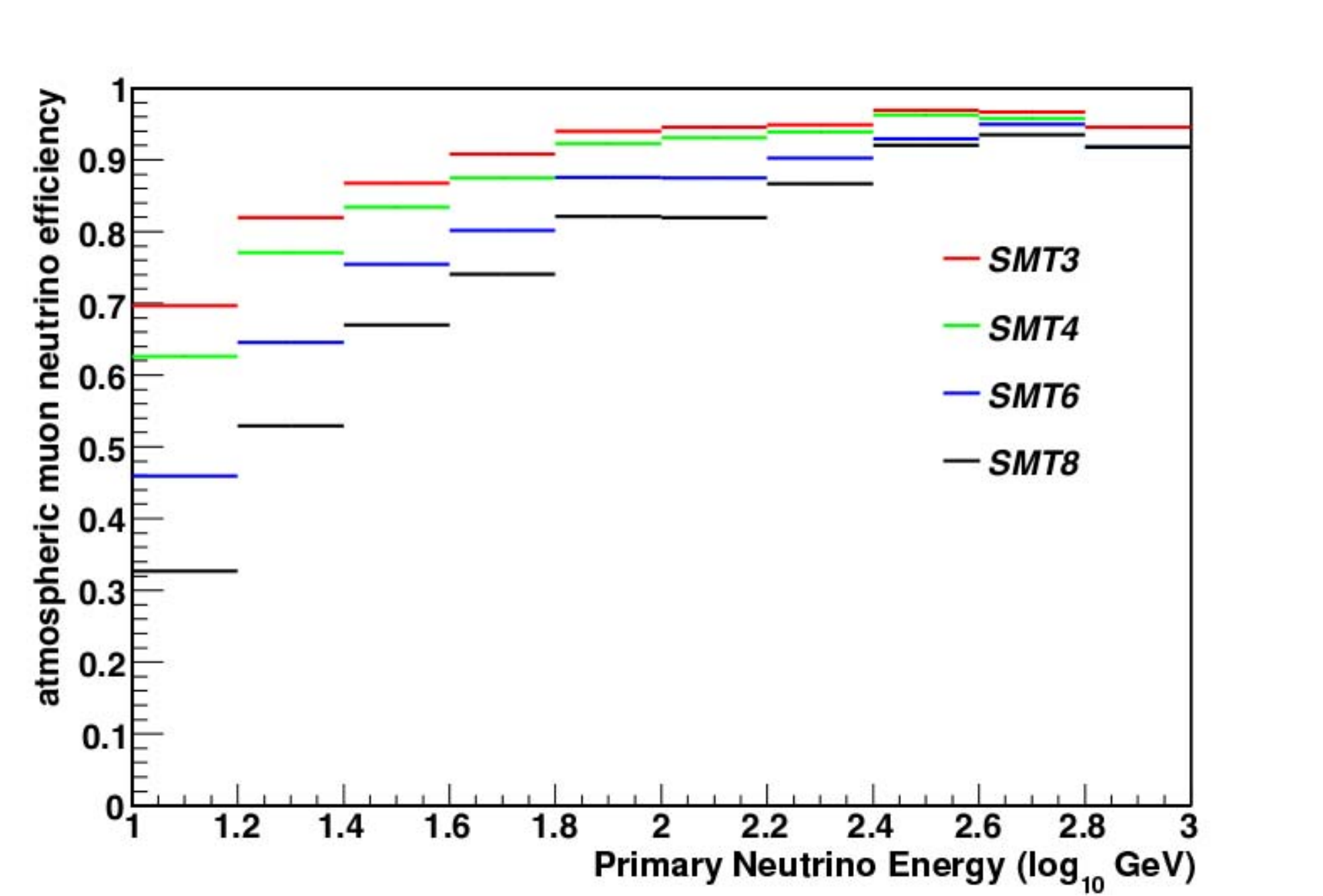}
   \end{center}
   \caption{Fraction of simulated atmospheric $\numu$ events
     satisfying different SMT trigger conditions.  The SMT3 condition is
     the loosest and hence admits the highest fraction of low energy
     events.}
   \label{fig:AtmNuTriggerEffVsE}
\end{figure}

Figure~\ref{fig:AtmNuTriggerEffVsE} shows the fraction of simulated
atmospheric $\numu$ events satisfying various DeepCore SMT trigger
conditions.  Since the trigger works only with HLC hits, an
event with only three HLC hits can have additional SLC hits.  These
additional hits will improve the reconstructability of such low
multiplicity events.  The minimum required multiplicity to reconstruct
a track is six, although depending on the distribution of these hits
along the strings, unavoidable ambiguities can arise.

\begin{figure} [h!]

  \centerline{
      \includegraphics[width=6cm]{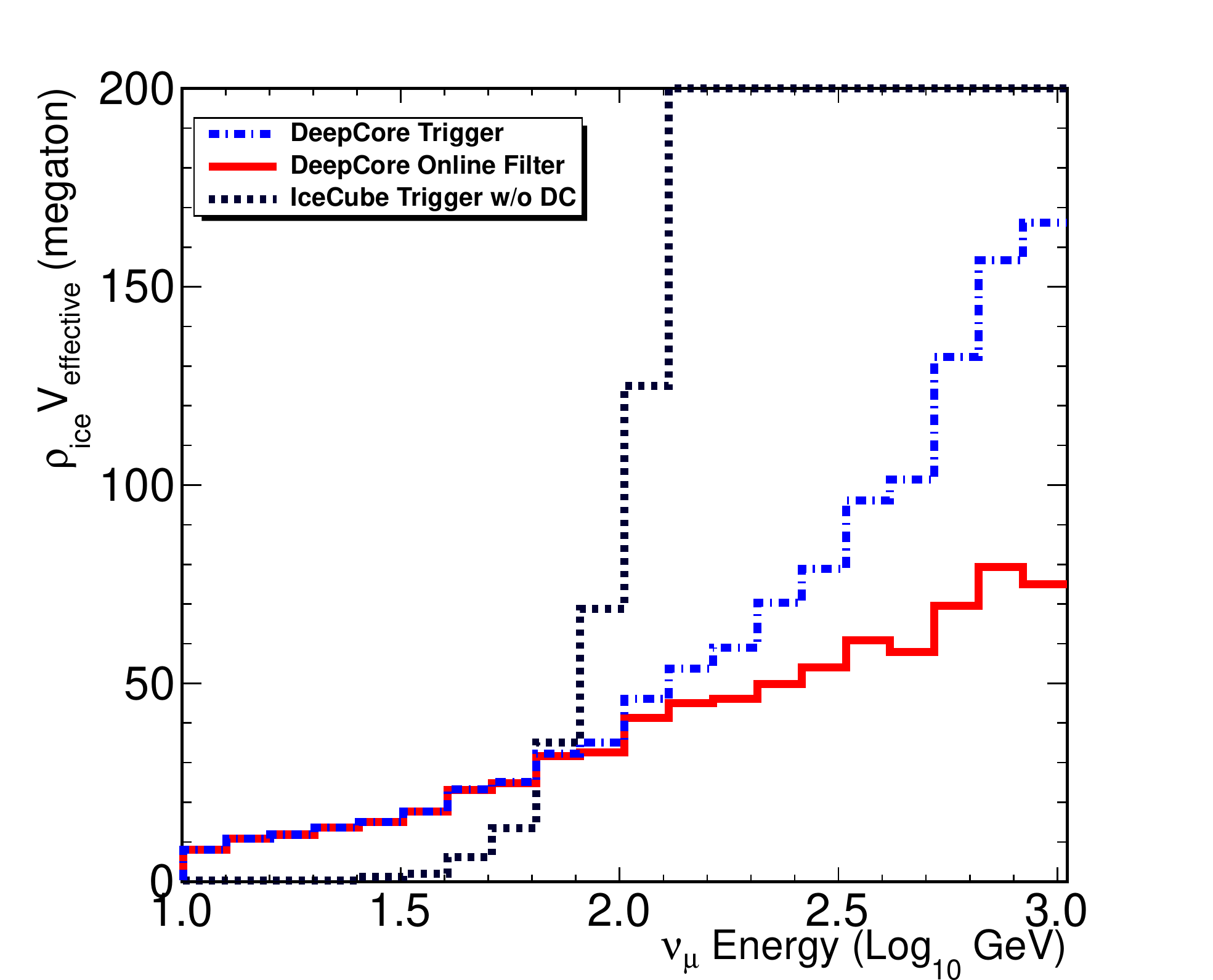}
      \hfill
      \includegraphics[width=6cm]{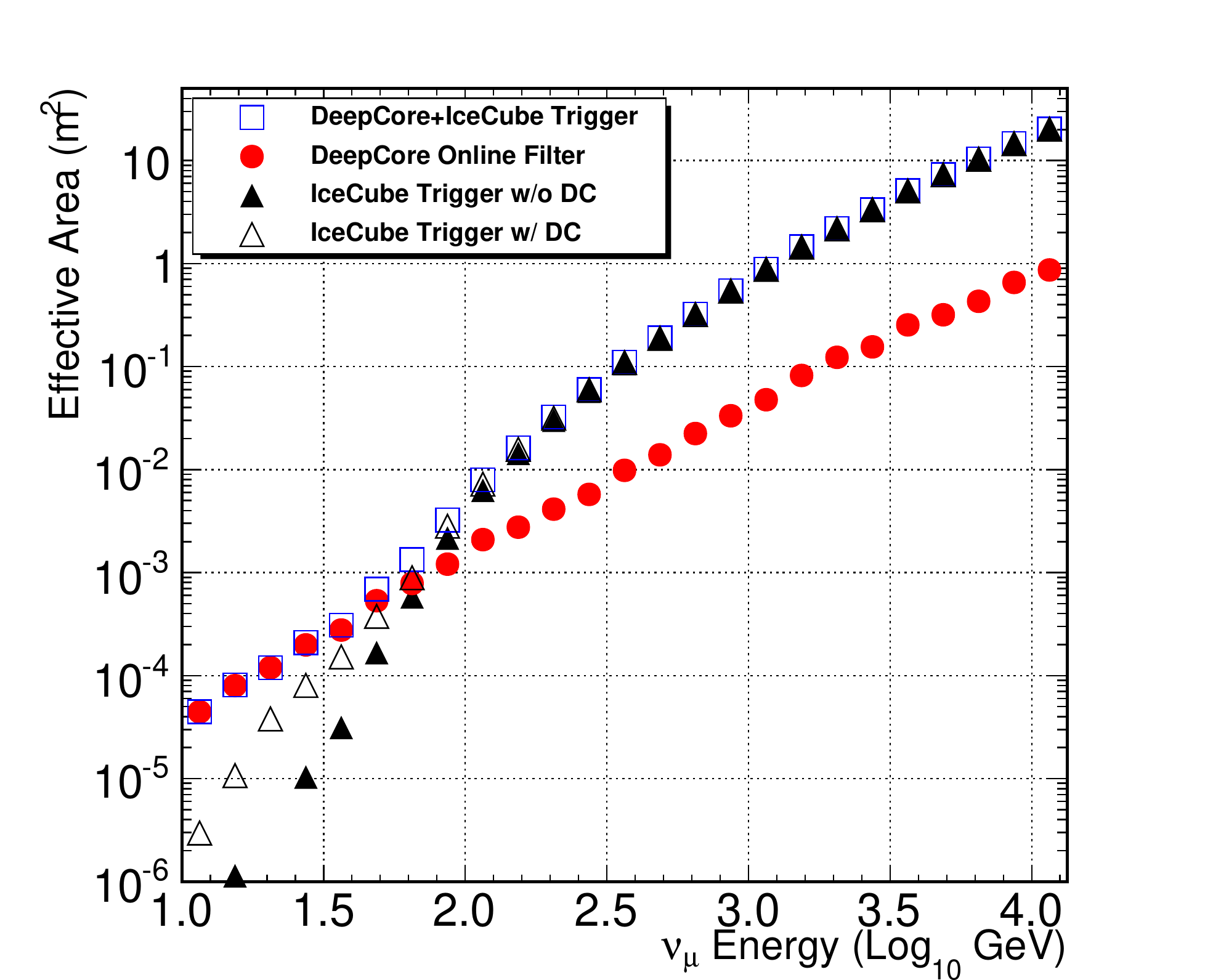}
      }
      \caption{Left: The expected DeepCore muon neutrino effective
        target mass, the product of the density of ice and the effective volume,
        after application of the SMT3 trigger (dash-dot line), after
        application of the online filter (solid line; see
        Sec.~\ref{subsec:Filtering}), after application of the SMT8
        trigger for the full 86-string IceCube detector (dotted line)
        and, to demonstrate the impact of DeepCore, after application
        of the SMT8 trigger for the IceCube detector simulated {\it
          without} DeepCore (dashed line).  The hadronic shower at the
        $\numu$ interaction vertex can contribute hits and play a role
        in the triggering and filtering.  High energy muon neutrinos
        that interact and produce a muon outside of the DeepCore
        volume are typically removed by the filter, eventually causing
        the solid line in the plot to turn over at energies above
        those shown.  From the DeepCore perspective, such events are
        indistinguishable from cosmic ray background, although many of
        them may be selected by the surrounding IceCube detector via
        other online filters.  Right: The expected DeepCore muon
        neutrino effective area after application of the SMT3 and SMT8
        triggers (open squares), after application of the online
        filter (solid circles; see Sec.~\ref{subsec:Filtering}), after
        application of the SMT8 trigger for the full 86-string IceCube
        detector (open triangles) and, to demonstrate the impact of
        DeepCore, after application of the SMT8 trigger for the
        IceCube detector simulated {\it without} DeepCore (solid
        triangles).  }
   \label{fig:VeffAeff_numu_withandwithoutDC}
\end{figure}

\begin{figure} [h!]
   \centerline{
      \includegraphics[width=6cm]{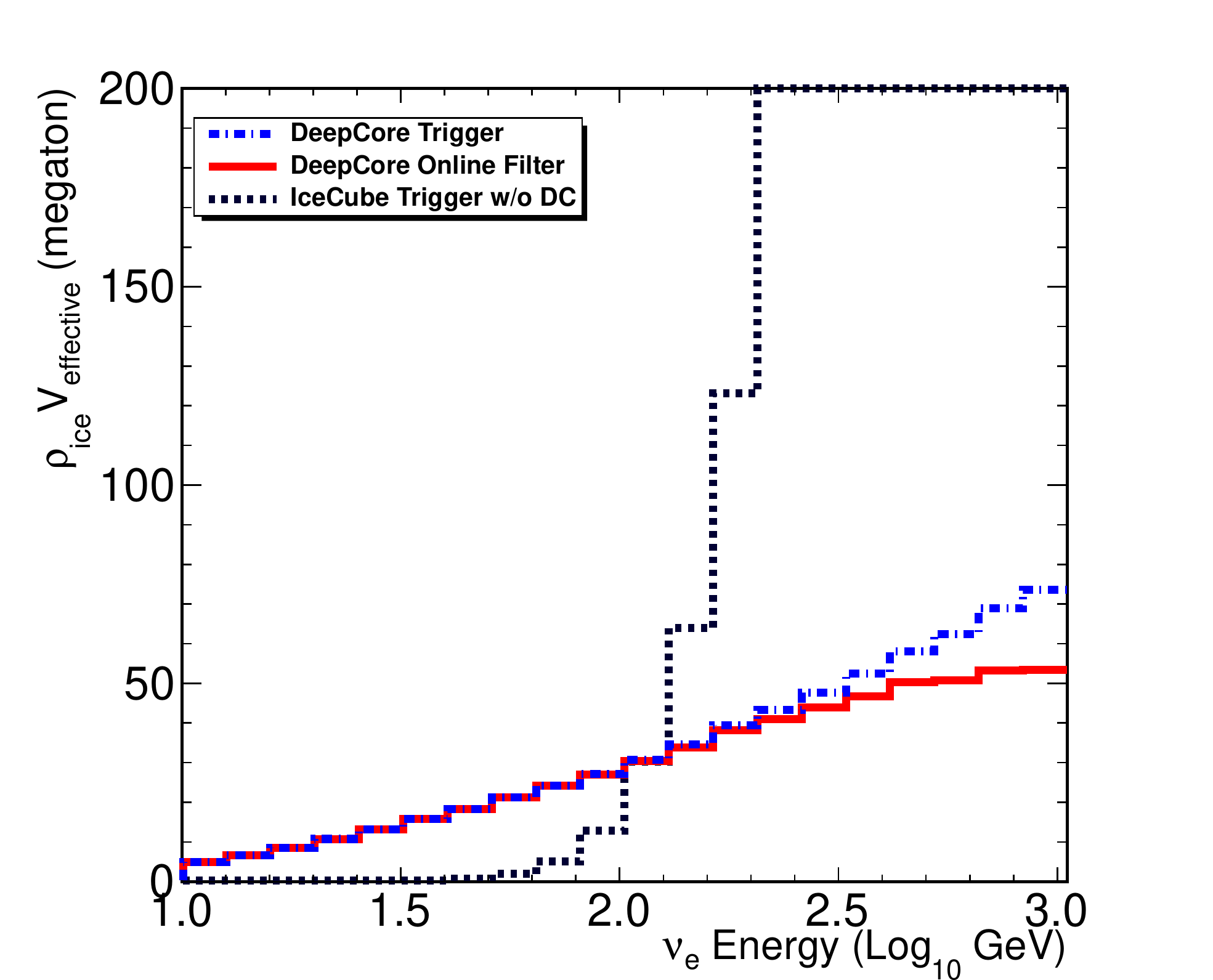}
      \hfill
      \includegraphics[width=6cm]{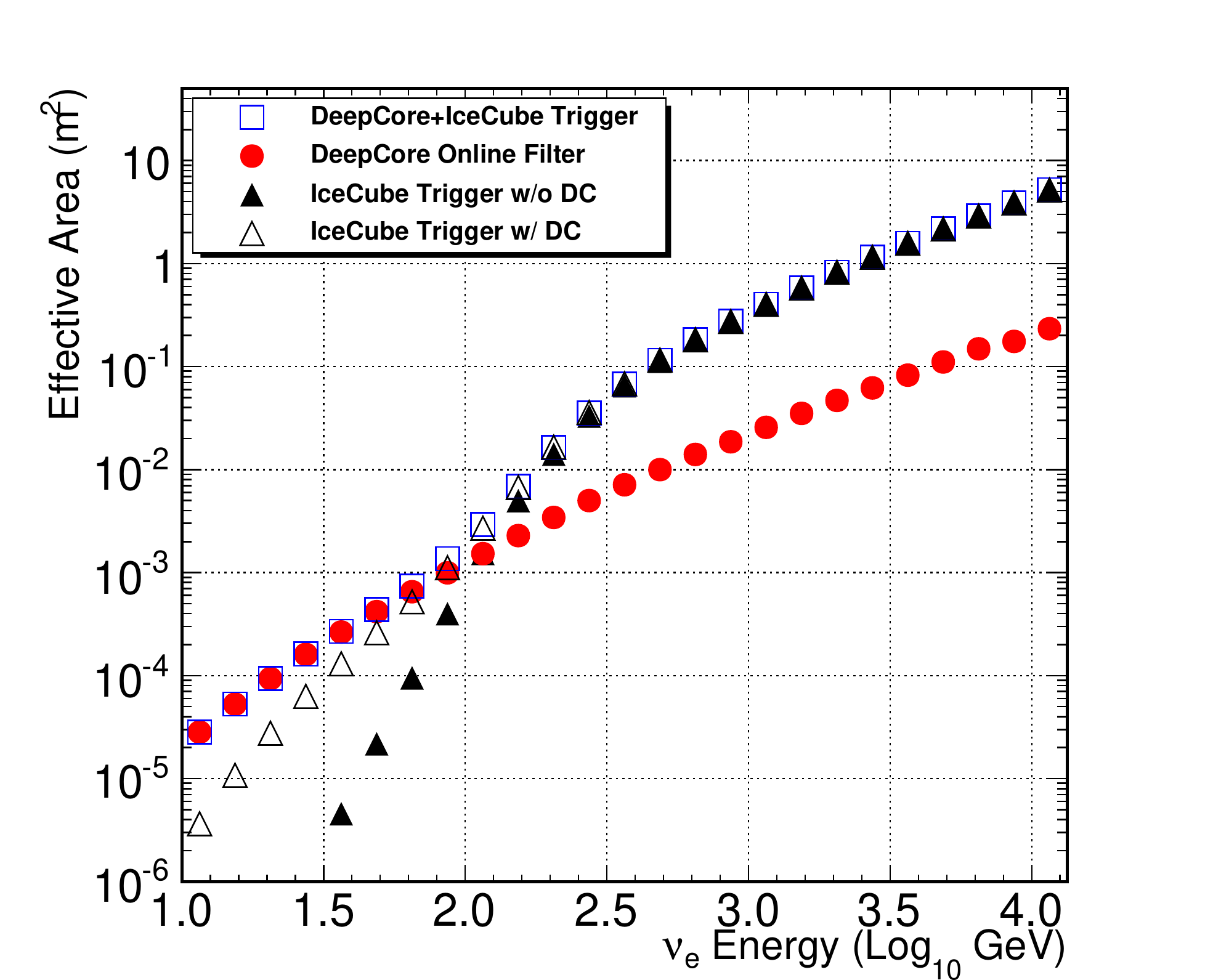}
   }
   \caption{Left: The expected DeepCore electron neutrino effective
     target mass, the product of the density of ice and the effective volume, after
     application of the SMT3 trigger (dash-dot line), after
     application of the online filter (solid line; see
     Sec.~\ref{subsec:Filtering}), after application of the SMT8
     trigger for the full 86-string IceCube detector (dotted line)
     and, to demonstrate the impact of DeepCore, after application of
     the SMT8 trigger for the IceCube detector simulated {\it without}
     DeepCore (dotted line).  High energy electron neutrinos that
     interact outside of the DeepCore volume are typically removed by
     the filter, causing the solid line in the plot to begin to turn
     over.  Many of these higher energy events may be captured by the
     surrounding IceCube detector via other online filters.  Right:
     The expected DeepCore electron neutrino effective area after
     application of the SMT3 trigger (open squares), after application
     of the online filter (solid circles; see
     Sec.~\ref{subsec:Filtering}), after application of the SMT8
     trigger for the full 86-string IceCube detector (open triangles),
     and, to demonstrate the impact of DeepCore, after application of
     the SMT8 trigger for the IceCube detector simulated {\it without}
     DeepCore (solid triangles).}
   \label{fig:VeffAeff_nue_withandwithoutDC}
\end{figure}

Figure~\ref{fig:VeffAeff_numu_withandwithoutDC} shows the effective
volume for muon neutrino events, 
\begin{equation}
   {\rm V}_{\rm eff} = {\rm V}_{\rm  gen} {\rm N}_{\rm trig}/{\rm N}_{\rm gen}, 
\end{equation}
where ${\rm V}_{\rm gen}$ is the volume in which the events were
generated and ${\rm N}_{\rm trig}$ and ${\rm N}_{\rm gen}$ the number
of events satisfying the trigger and the number generated,
respectively.  Figure~\ref{fig:VeffAeff_numu_withandwithoutDC} also
shows the effective area, ${\rm A}_{\rm eff}$, for muon neutrino
events satisfying the SMT3 trigger condition as a function of energy.
The definition for \Aeff parallels that of \Veff.
Figure~\ref{fig:VeffAeff_nue_withandwithoutDC} shows the effective
volume and area for electron neutrinos.  To further demonstrate the
impact of DeepCore for low energy physics, these figures also show the
same quantities as described above but with DeepCore artificially
removed in the simulation of the response of the full detector.

Figures~\ref{fig:VeffAeff_numu_withandwithoutDC}-\ref{fig:VeffAeff_nue_withandwithoutDC}
were all simulated using the 86-string configuration of IceCube
(``IC86'') that includes the 15 strings of DeepCore.

More specialized triggers that use data from a combination of IceCube
and DeepCore modules are under development to enhance sensitivity to
particular signals, such as neutrino signals from solar WIMP
annihilations.  These specialized triggers include those that can be
implemented in the trigger software, sensitive to specific event
topologies, and those that can be implemented using new hardware
operating in conjunction with the existing DAQ system.  These triggers
will be described in a future publication.

%% file: Filtering.tex
Due to its location at the South Pole, real-time communication with
IceCube from the northern hemisphere can only be provided by
geosynchronous satellites with limited transmission bandwidth.  Since
the background flux of cosmic-ray muons is about $10^6$ times larger
than the flux of atmospheric neutrinos in the full IceCube array,
individual analyses employ software ``filters'' that reduce the size
of the data sample by selecting likely signal events and removing
likely background events.  The filtered subset of the triggered data
stream is transmitted daily to storage facilities in the north, where
more sophisticated reconstruction algorithms are applied to the data.
All events are written to portable storage media at the South Pole and
transported north for archival storage at the end of each austral
winter.

With DeepCore fully deployed and surrounded by standard IceCube
strings, a new filter taking full advantage of the vetoing
capabilities of the surrounding strings is being used.  This new
filter is distinct from the standard IceCube filters designed to
enrich potential signal relative to background for a variety of event
topologies.  The design and performance of this filter are described
below.  Additional and more sophisticated vetoing algorithms will be
applied to filtered data offline in the north.  The offline veto
algorithms are still under development.  Since the cosmic-ray muon
flux is attenuated by about an order of magnitude relative to that of
IceCube by virtue of DeepCore's greater average overburden, the
overall goal of the trigger, online filter and offline veto is to
attain a cosmic-ray muon rejection factor of at least $10^5$.  At the
same time, we aim to maintain a signal efficiency of well over 50\%
for contained and partially contained neutrino-induced tracks and
showers down to $\Enu \gsim 10$~GeV.

The online filter is used to search for HLC hits in the ``veto''
region external to DeepCore's fiducial volume that are consistent with
the presence of a downward-going muon.  The online filter provides an
estimate of the ``center of gravity'' (COG) and time of the event
within DeepCore by calculating the average position $r$ and time $t$
of the DOM hits in DeepCore.  The initial COG estimate is then refined by
using the average position, $r^\prime$, of the subset of those hits
with times within one standard deviation of the average time.  The
initial time estimate is refined by using the average of the
``corrected'' hit times, $t^\prime$.  Corrected hit times are
determined by subtracting from the time of each hit the time that
unscattered light would require to travel from the COG at $r^\prime$.

\begin{figure}[!t]
  \centering
  \includegraphics[width=2.5in]{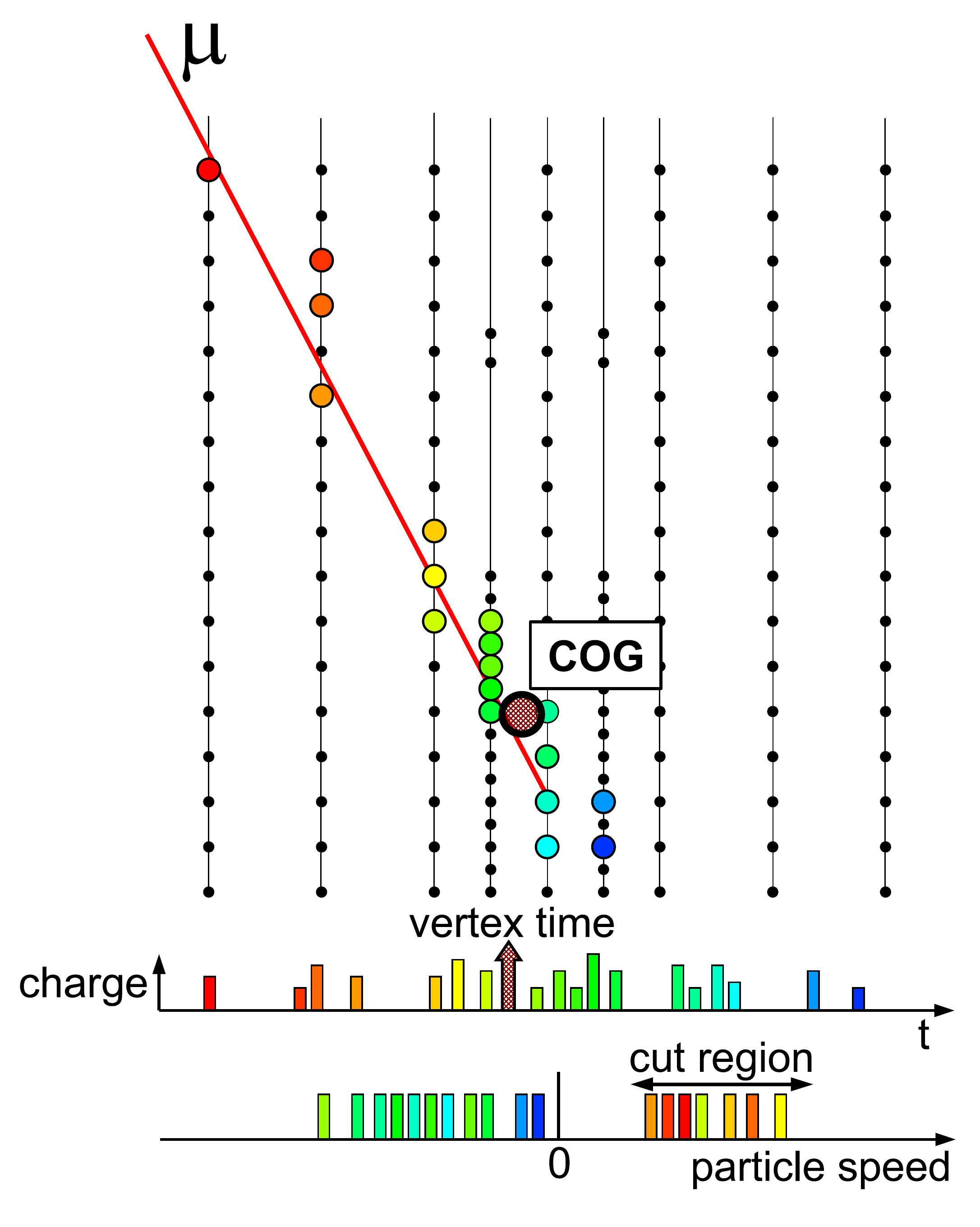}
  \caption{A simulated downward-going muon event that would be vetoed
    by the algorithm described in the text.  The vertical lines
    represent strings and the small black circles represent individual
    DOMs.  The larger circles at DOM positions represent hits.  The
    earliest hits are in red and the latest in violet, with hit times
    in between following the colors of the rainbow.  The center of
    gravity of the hits in the DeepCore volume is labelled COG.  The
    hits in the upper left hand side, colored red and orange, are the
    early hits associated with the muon's entry point into the
    detector fiducial volume, and these hits have associated
    ``particle speeds'' consistent with speed of light travel between
    the hit and the COG, and therefore are consistent with having been
    produced by a muon.  These hits enter the ``cut region'' in
    particle speed, shown at the bottom of the figure, and the event
    is vetoed on this basis. }
  \label{fig:FilterDiagram}
\end{figure}

With this refined COG estimate, the online filter is used to calculate the
speed of a hypothetical particle traveling from each HLC hit in the
surrounding IceCube volume (used as a veto region) to the COG.
Events that have at least one hit with a speed consistent with $v=c$,
where the speed $v = (r^\prime-r_{\rm DOM})/(t^\prime-t_{\rm DOM})$,
are rejected.  This algorithm is depicted graphically in
Fig.~\ref{fig:FilterDiagram}.

\begin{figure}[!t]
  \centering
  \includegraphics[width=2.5in]{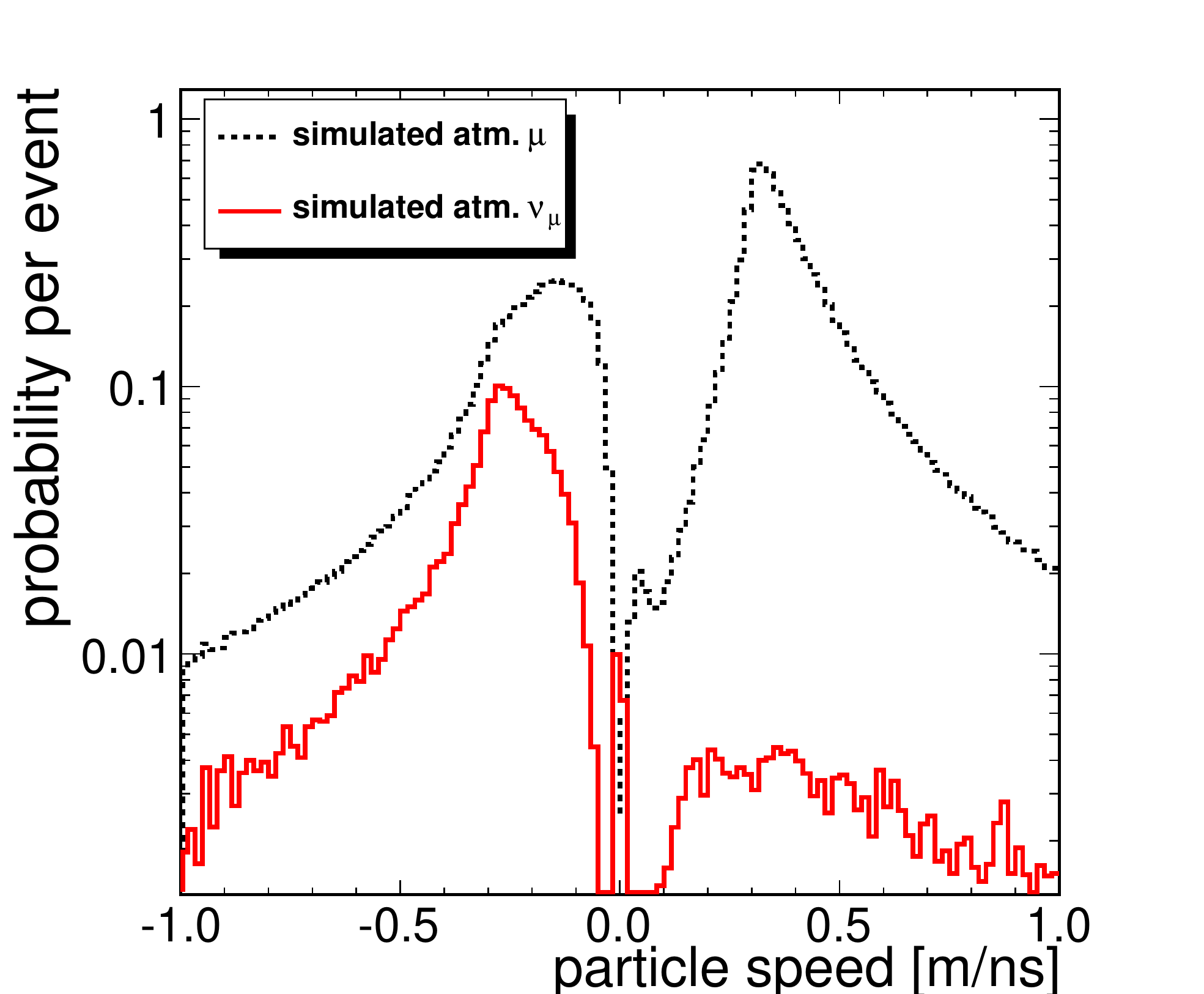}
  \caption{Particle speed probabilities per event for simulated muons
    from cosmic-ray interactions (black dashed line) and simulated muons
    from atmospheric neutrinos inside DeepCore (red solid line).  The
    speed is defined to be positive if the hit occurred before the COG
    time (see text) and negative if it appeared after.  Hits in the
    veto region are generally expected to have a speed close to $c
    \simeq +0.3$~m/ns.  Smaller speeds occur for light delayed by
    scattering. Larger speeds are in principle acausal, but since the
    COG time represents the start of a DeepCore event, whereas the COG
    position defines its center, the particle speeds for early hits
    are slightly overestimated.  Events with a hit within a particle
    speed window between +0.25 and +0.4~m/ns are rejected.}
  \label{fig:ParticleSpeedProbabilities}
\end{figure}

Figure~\ref{fig:ParticleSpeedProbabilities} shows the distribution of
hypothetical particle speeds per event. The dotted curve depicts the
simulated muon background from cosmic-ray air showers using CORSIKA
and the solid curve the atmospheric neutrino signal~\cite{Bartol}.
The atmospheric neutrino events are required to have an interaction
vertex inside the DeepCore fiducial volume, as determined from Monte
Carlo truth information.  The peak for the simulated cosmic-ray muons
is slightly above +0.3~m/ns while muons induced by neutrinos in
DeepCore mainly give hits with negative particle speeds.  Negative
speeds indicate that the hypothetical particle traveled outward from
the fiducial volume into the veto volume.  The peak at positive speeds
close to zero is mainly due to early scattered light.  By rejecting
events with one hit within a particle speed window between +0.25 and
+0.4~m/ns we achieve an overall background rejection of roughly
$8\cdot10^{-3}$.

\begin{figure}[!t]
  \centering
  \includegraphics[width=2.5in]{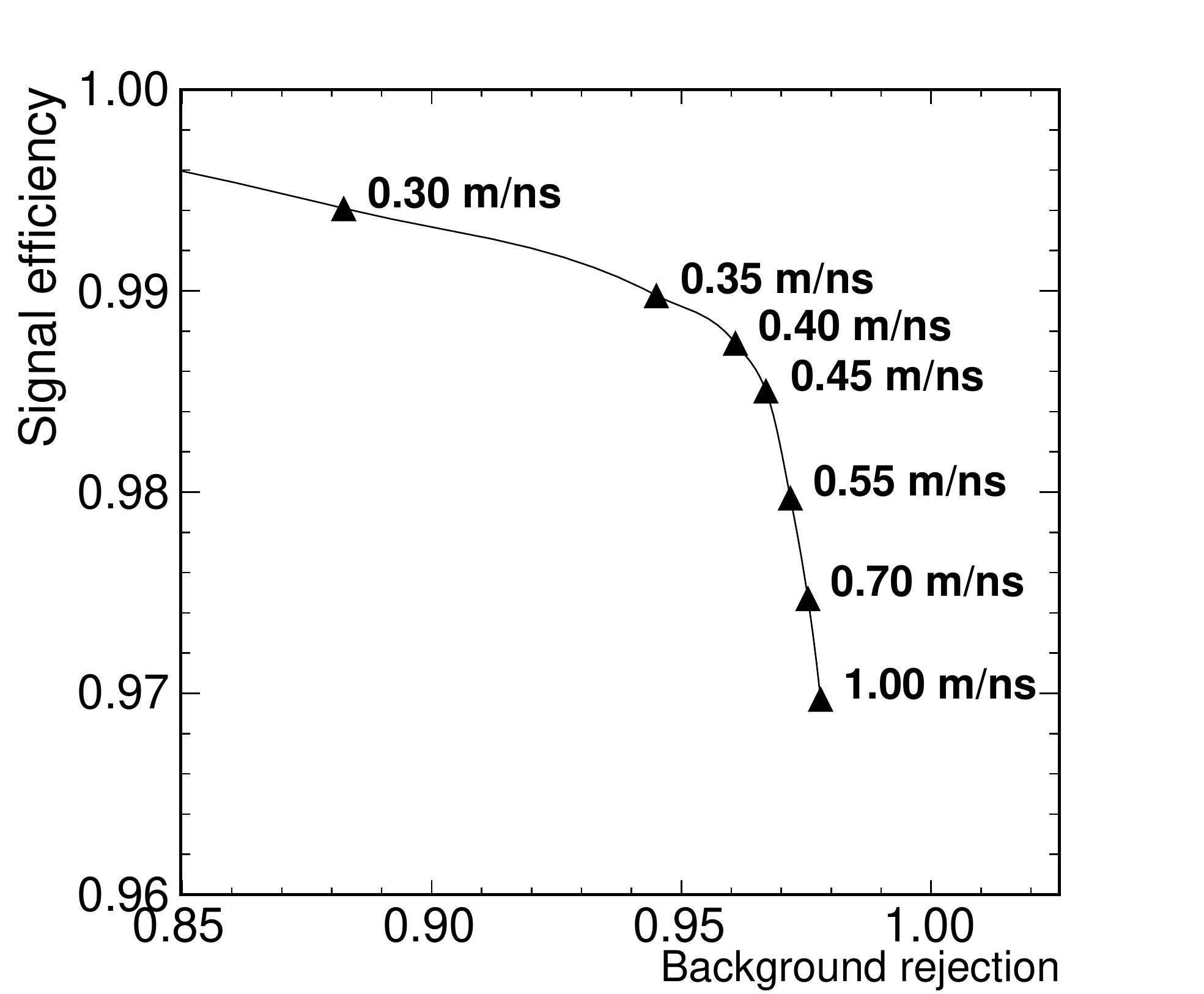}
  \caption{Signal efficiency as a function of background rejection for
    events having one or more hits with a particle speed (see text)
    between +0.25~m/ns and a variety of upper values, ranging from
    +0.35~m/ns to +1.0~m/ns, as indicated in the figure.  Upper values
    higher than about +0.4~m/ns result in greater signal loss without
    significant additional background rejection.}
  \label{fig:VetoWindowCutStop}
\end{figure}

Figure~\ref{fig:VetoWindowCutStop} shows the signal efficiency
vs. background rejection for events that have one or more hits with a
particle speed between +0.25~m/ns and a range of maximum speeds from
+0.3 to +1.0~m/ns.  As the value of the maximum speed increases,
signal efficiency decreases more quickly than background rejection
increases.  Also taking satellite bandwidth limitations into
consideration, we set the maximum allowable speed to +0.4~m/ns.
Similarly, varying the minimum speed while holding the maximum speed
fixed at +0.4~m/ns, we set the minimum allowable speed to +0.25~m/ns.

As we have enough bandwidth capacity we can effort to send the data
with 96\% background rejection and keep highest possible signal
efficiency. More strict selection criteria start to decrease the
signal efficiency, so that we choose 0.4m/ns as selection cut.

The background rejection and signal detection rates of the online
filter are compiled in Table~\ref{tab:FilterPerformance}. For IceCube
in its 79-string configuration, the online filter passed data at about
4~GB/day, about 5\% of the available satellite bandwidth allocated to
IceCube.

\begin{table*}[th]

  \caption{Data and signal passing rates (in Hz) after
    application of the DeepCore trigger and online filter.  
    In anticipation of future selection criteria that will require
    reconstructed interaction vertices to be well contained in the DeepCore fiducial volume, 
    only atmospheric neutrinos interacting inside the detector fiducial volume 
    (a cylinder of radius 200~m and height 350~m) were simulated.
    The online filter has negligible impact on simulated signal events
    while reducing the data, which is dominated by downward-going muons, by about a factor of 
    15 relative to the SMT3 trigger.  The data used here came from the 79-string configuration 
    of IceCube (``IC79'') while the simulated signal used the 86-string configuration of 
    IceCube (``IC86'').  Numbers for the data in IC86
    running will be marginally higher than those shown here.}

  \label{tab:FilterPerformance}
  \centering
  \begin{tabular}{|l|c|c|c|}
    \hline
    Rates (Hz) 
       & Data (IC79)             & Atm. $\numu$ (IC86)      & Atm. $\nue$ (IC86)         \\
       & (annual                 & (Bartol~\cite{Bartol})   & (Bartol~\cite{Bartol})     \\
       & average)                & ($\uparrow$: upward)     & ($\uparrow$: upward)       \\
       &                         & ($\downarrow$: downward) & ($\downarrow$: downward)   \\  \hline 
    %-------------------------------------------------------------------------------------------
    All IceCube 
       & 1900                    &                          &                            \\ 
    Triggers       
       &                         &                          &                            \\  \hline

    DeepCore 
       & 185 (9.7\%)             & 3.59$\cdot 10^{-3}$ (100\%)& 0.793$\cdot 10^{-3}$ (100\%)\\
    SMT3 
       &                         & ($\uparrow$: 1.54$\cdot 10^{-3}$)
                                                            & ($\uparrow$: 0.411$\cdot 10^{-3}$) \\
    Trigger                                                                              
       &                         & ($\downarrow$: 2.05$\cdot 10^{-3}$)
                                                            & ($\downarrow$: 0.382$\cdot 10^{-3}$)
                                                                                          \\  \hline

    DeepCore 
       &  17.5 (0.9\%)           & 3.57$\cdot 10^{-3}$ (99.4\%) & 0.789$\cdot 10^{-3}$ (99.5\%)\\
    Online
       &                         & ($\uparrow$: 1.53$\cdot 10^{-3}$)
                                                            & ($\uparrow$: 0.409$\cdot 10^{-3}$)\\
    Filter                                                                                   
       &                         & ($\downarrow$: 2.04$\cdot 10^{-3}$)
                                                            & ($\downarrow$: 0.380$\cdot 10^{-3}$)
                                                                                            \\  \hline

%    SMT8 
%       &           &                      &                        &                       & \\
%    Trigger
%       &           &                      &                        &                       & \\
%    \hline
  \end{tabular}
\end{table*}

The effect of the online filter on the muon neutrino effective volume
and effective area is shown in Fig.~\ref{fig:VeffAeff_numu_withandwithoutDC}.
Its effect for electron neutrinos is shown in
Fig.~\ref{fig:VeffAeff_nue_withandwithoutDC}.

\begin{figure} [h!]

  \centerline{
      \includegraphics[width=6cm]{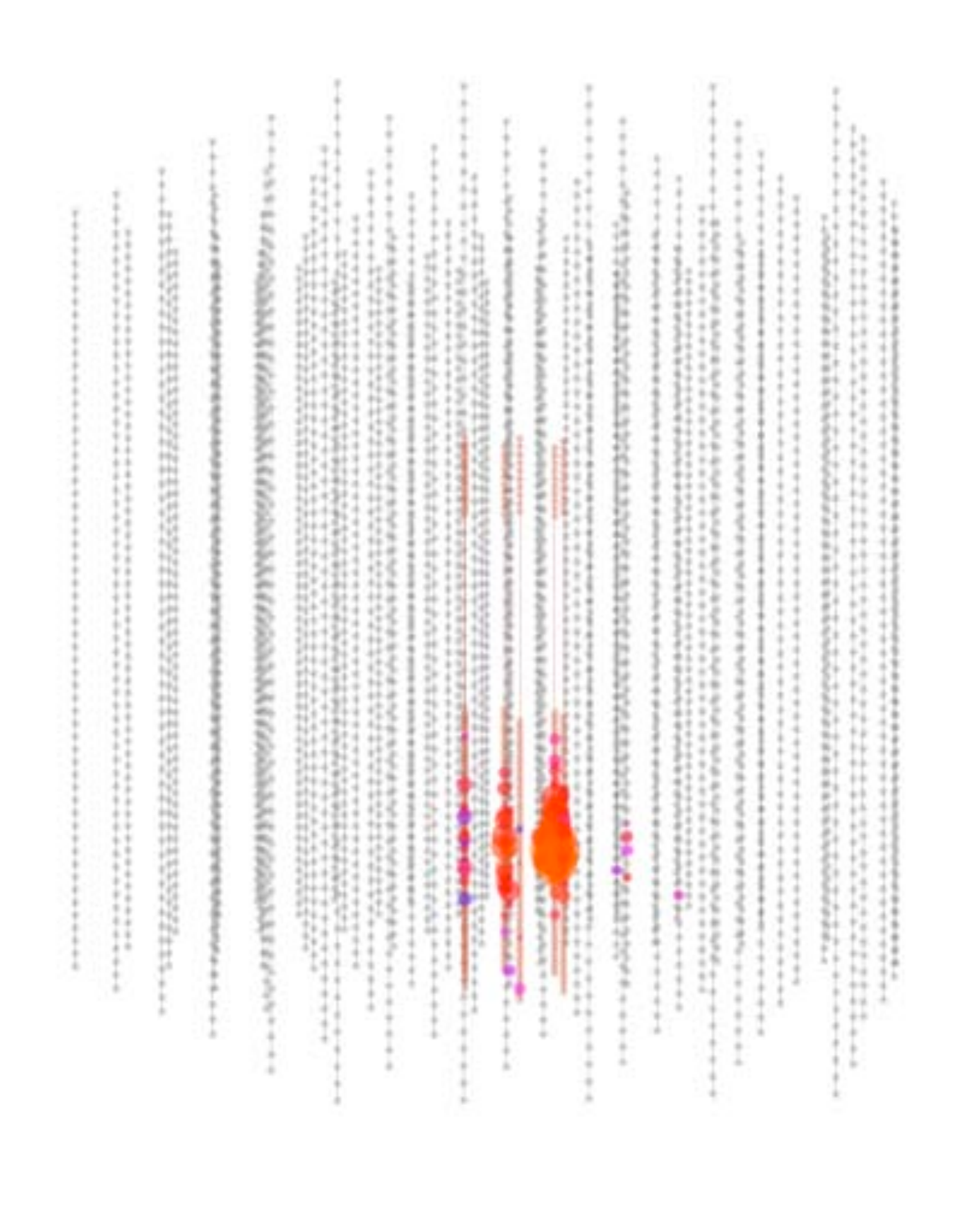}
      \hfill
      \includegraphics[width=6cm]{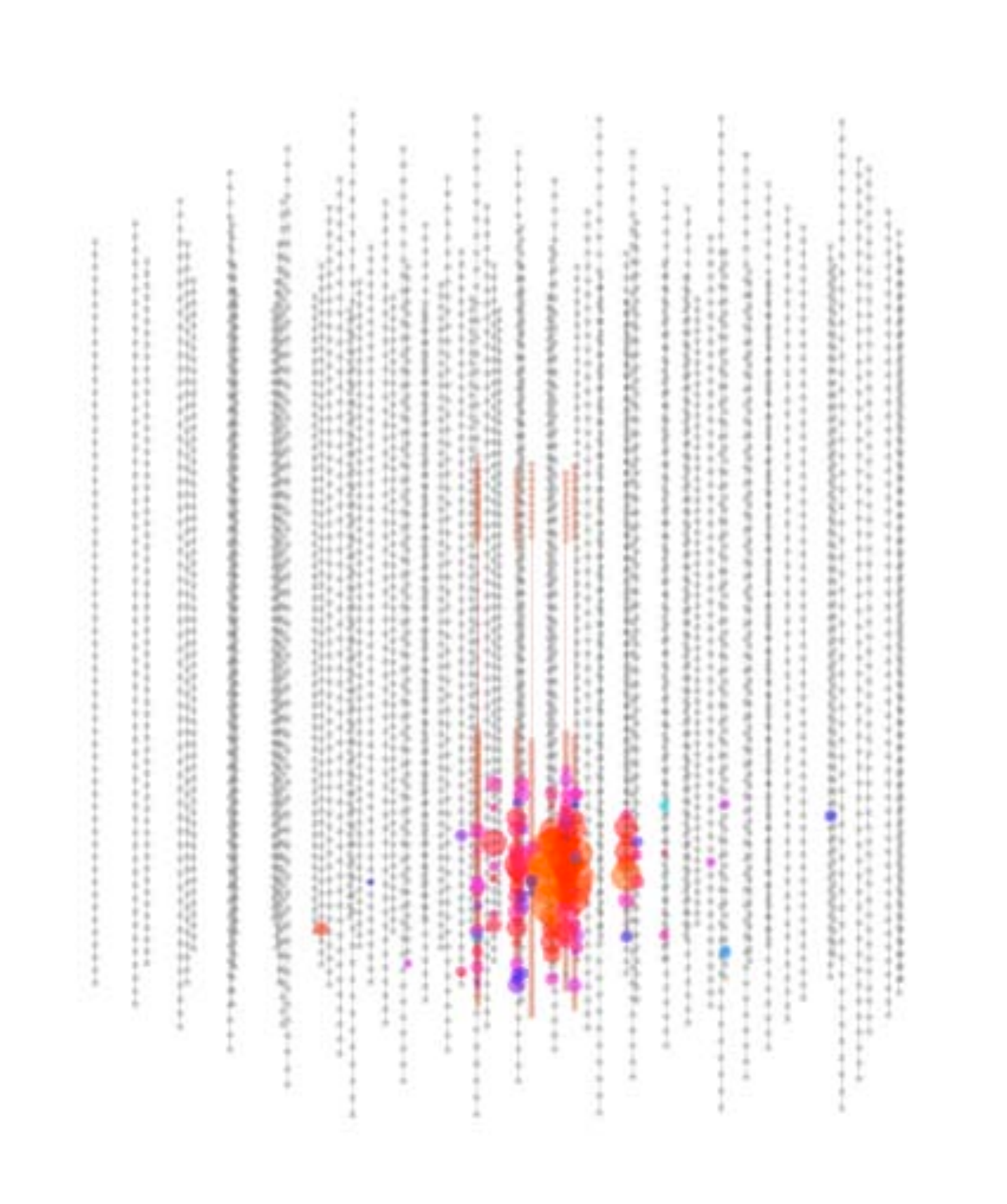}
      }
      \caption{Two candidate shower events induced by atmospheric
        neutrino interactions in the DeepCore fiducial volume.  The
        event energies are on the order of $10^2$~GeV (left) and
        $10^3$~GeV (right).  Each black point represents a single DOM.
        Points with superimposed colored circles represent DOMs that
        received light associated with the event, with the size of the
        circle proportional to the amount of light, and the color
        indicating the relative time of arrival of the first photon
        detected by that DOM, with red earliest and violet latest
        (following the colors of the rainbow).  }
   \label{fig:DeepCoreCascade06and07}
\end{figure}

The analysis of the first year of IceCube DeepCore data is underway.
One of the first analyses nearing completion is a measurement of
hadronic and electromagetic showers induced by atmospheric neutrinos
in the DeepCore fiducial volume~\cite{DeepCoreIC79CascadesICRC}.  In
Fig.~\ref{fig:DeepCoreCascade06and07}, two candidate shower events
with energies on the order of $10^2$~GeV (left) and $10^3$~GeV (right) are
shown.  These events were extracted from the data after application of
the triggering and filtering criteria described above, along with a
variety of additional, more sophisticated selection criteria.

\clearpage

%% file: Conclusions.tex
The IceCube DeepCore sub-array has been deployed and is actively
taking analysis-quality physics data.  It was designed to be sensitive
to neutrino energies as low as about 10~GeV, over an order of
magnitude lower than the original goal for IceCube.  Situated in the
very clear ice more than 2100~m below the surface, outfitted with new
high quantum efficiency photomultiplier tubes, and deployed on a very
close spacing, DeepCore is operating as anticipated and is expected to
reach its low energy goal. We have successfully implemented simple and
robust online algorithms that reduce the background level by over two
orders of magnitude while retaining most of the expected neutrino
signal.  More sophisticated algorithms, run offline in the north,
should allow us to further reduce the background to a level comparable
to the atmospheric neutrino flux.  This should give DeepCore sensitivity
to a wide range of exciting physics, from low mass solar WIMP
annihilations and atmospheric neutrino oscillations to soft-spectrum
point sources of neutrinos in the southern sky and exotic physics such
as slow-moving monopoles.

%% file: Acknowledgements.tex
We acknowledge the support from the following agencies: U.S. National
Science Foundation-Office of Polar Programs, U.S. National Science
Foundation-Physics Division, University of Wisconsin Alumni Research
Foundation, the Grid Laboratory Of Wisconsin (GLOW) grid
infrastructure at the University of Wisconsin - Madison, the Open
Science Grid (OSG) grid infrastructure; U.S. Department of Energy, and
National Energy Research Scientific Computing Center, the Louisiana
Optical Network Initiative (LONI) grid computing resources; National
Science and Engineering Research Council of Canada; Swedish Research
Council, Swedish Polar Research Secretariat, Swedish National
Infrastructure for Computing (SNIC), and Knut and Alice Wallenberg
Foundation, Sweden; German Ministry for Education and Research (BMBF),
Deutsche Forschungsgemeinschaft (DFG), Research Department of Plasmas
with Complex Interactions (Bochum), Germany; Fund for Scientific
Research (FNRS-FWO), FWO Odysseus programme, Flanders Institute to
encourage scientific and technological research in industry (IWT),
Belgian Federal Science Policy Office (Belspo); University of Oxford,
United Kingdom; Marsden Fund, New Zealand; Japan Society for Promotion
of Science (JSPS); the Swiss National Science Foundation (SNSF),
Switzerland; A. Groß acknowledges support by the EU Marie Curie OIF
Program; J. P. Rodrigues acknowledges support by the Capes Foundation,
Ministry of Education of Brazil.

%% file: Main-Els.bbl
\begin{thebibliography}{10}
\expandafter\ifx\csname url\endcsname\relax
  \def\url#1{\texttt{#1}}\fi
\expandafter\ifx\csname urlprefix\endcsname\relax\def\urlprefix{URL }\fi
\expandafter\ifx\csname href\endcsname\relax
  \def\href#1#2{#2} \def\path#1{#1}\fi

\bibitem{IceCube}
A.~Achterberg, et~al., First {Y}ear {P}erformance of the {I}ce{C}ube {N}eutrino
  {T}elescope, Astropart. Phys. 26 (2006) 155--173.

\bibitem{IC22WIMPs}
R.~Abbasi, et~al., Phys. Rev. Lett. 102 (2009) 201302.

\bibitem{WIMPs_GC}
M.~Bissok, D.~Boersma, J.-P. Huelss, C.~Rott, Search for {D}ark {M}atter in the
  {M}ilky {W}ay with {I}ce{C}ube, to be published in Proceedings of the 32nd
  International Cosmic Ray Conference, Beijing.

\bibitem{WIMPS_GH}
R.~Abbasi, et~al., Phys. Rev. D 84 (2011) 022004.

\bibitem{Mena:2008rh}
O.~Mena, I.~Mocioiu, S.~Razzaque, Phys. Rev. D 78 (2008) 093003.

\bibitem{2010PhRvD81h3011T}
I.~{Taboada}, Multiflavor and {M}ultiband {O}bservations of {N}eutrinos from
  {C}ore {C}ollapse {S}upernovae, Phys. Rev. D 81~(8) (2010) 083011.

\bibitem{SN_ICRC}
V.~Baum, L.~Demir\"ors, L.~K\"opke, M.~Ribordy, et~al., {Supernova Detection
  with {I}ce{C}ube and Beyond}, to be published in Proceedings of the 32nd
  International Cosmic Ray Conference, Beijing.

\bibitem{SN_ArXiv}
L.~Demir\"ors, M.~Ribordy, M.~Salathe, Novel {T}echnique for {S}upernova
  {D}etection with {I}ce{C}ube\href {http://arxiv.org/abs/1106.1937v2}
  {\path{arXiv:1106.1937v2}}.

\bibitem{staus}
I.~F.~M. Albuquerque, G.~Burdman, Z.~Chacko, Direct {D}etection of
  {S}upersymmetric {P}articles in {N}eutrino {T}elescopes, Phys. Rev. D 75~(3)
  (2007) 035006.

\bibitem{SUSYKaluzaKleinMultipleScattering}
I.~F.~M. Albuquerque, S.~R. Klein, {Supersymmetric and Kaluza-Klein Particles
  Multiple Scattering in the Earth}, Phys. Rev. D 80~(1) (2009) 015015.

\bibitem{LowEGRBNus}
J.~Bahcall, P.~M\'esz\'aros, Phys. Rev. Lett. 85 (2000) 1362.

\bibitem{IceCubeDOMs}
R.~Abbasi, et~al., Calibration and {C}haracterization of the {I}ce{C}ube
  {P}hotomultiplier {T}ube, Nucl. Inst. Meth. A 618~(1-3) (2010) 139--152.

\bibitem{IceCubeDAQ}
R.~Abbasi, et~al., The {I}ce{C}ube {D}ata {A}cquisition {S}ystem: {S}ignal
  {C}apture, {D}igitization, and {T}imestamping, Nucl. Inst. and Meth. A601
  (2009) 294--316.

\bibitem{icepaper}
M.~Ackermann, et~al., Optical {P}roperties of {D}eep {G}lacial {I}ce at the
  {S}outh {P}ole, J. Geophys. Res. 111 (2006) D13203.

\bibitem{IceTemperatureProfile}
P.~Price, et~al., Temperature {P}rofile for {G}lacial {I}ce at the {S}outh
  {P}ole: {I}mplications for {L}ife in a {N}earby {S}ubglacial {L}ake,
  Proceedings of the National Academy of Sciences 99~(12) (2002) 7844--7847.

\bibitem{IceCubeDeployment}
A.~Karle, {{I}ce{C}ube: {C}onstruction {S}tatus and {F}irst {R}esults}, Nucl.
  Inst. Meth. A~(604) (2009) S46--S52.

\bibitem{CORSIKA}
D.~Heck, et~al., {CORSIKA}: {A} {M}onte {C}arlo {C}ode to {S}imulate
  {E}xtensive {A}ir {S}howers, Tech. Rep. FZKA 6019 (1998) 1--90.

\bibitem{ANIS}
A.~Gazizov, M.~Kowalski, {ANIS}: {H}igh {E}nergy {N}eutrino {G}enerator for
  {N}eutrino {T}elescopes, Comput. Phys. Commun. 172 (2005) 203--213.

\bibitem{GENIE}
C.~Andreopoulos, et~al., The {GENIE} {N}eutrino {M}onte {C}arlo {G}enerator,
  Nucl. Instrum. Meth. A614 (2010) 87--104.

\bibitem{MMC}
D.~Chirkin, W.~Rhode, {P}ropagating {L}eptons {T}hrough {M}atter with {M}uon
  {M}onte {C}arlo ({MMC})\href {http://arxiv.org/abs/hep-ph/0407075v2}
  {\path{arXiv:hep-ph/0407075v2}}.

\bibitem{PHOTONICS}
J.~Lundberg, et~al., Light {T}racking {T}hrough {I}ce and {W}ater -
  {S}cattering and {A}bsorption in {H}eterogeneous {M}edia with {PHOTONICS},
  Nucl. Inst. Meth. A581 (2007) 619--631.

\bibitem{Bartol}
G.~D. Barr, et~al., Phys. Rev. D 70 (2004) 023006.

\bibitem{DeepCoreIC79CascadesICRC}
C.~H. Ha, D.~J. Koskinen, et~al., Observation of {A}tmospheric
  {N}eutrino-{I}nduced {C}ascades in {I}ce{C}ube-{D}eep{C}ore, to be published
  in Proceedings of the 32nd International Cosmic Ray Conference, Beijing.

\end{thebibliography}
